\documentclass[11pt]{article}

\usepackage{amsfonts}
\usepackage{amsmath}
\usepackage{empheq}
\usepackage{changepage}
\usepackage{slashed}
\usepackage{mathtools}
\usepackage{putex}

\usepackage[normalem]{ulem}
\usepackage{etoolbox}
\usepackage{cancel}

\newcommand{\reef}[1]{(\ref{#1})}

\def\be{\begin{equation}}
\def\ee{\end{equation}}
\def\bea{\begin{eqnarray}}
\def\eea{\end{eqnarray}}
\def\ba{\begin{array}}
\def\ea{\end{array}}
\def\bd{\begin{displaymath}}
\def\ed{\end{displaymath}}

\def\tr{{\rm tr}}

\def\a{\alpha}
\def\b{\beta}

\def\d{\delta}
  
\def\g{\gamma}

\def\l{\lambda}
\def\m{\mu}

\def\s{\sigma}                                   

\def\D{\Delta}

\def\G{\Gamma}

\def\O{\Omega}

\def\pa{\partial}                              
\def\>{\rangle} 
\def\<{\langle} 
\def\Dsl{D \hskip-.6em \raise1pt\hbox{$ / $ } }
\def\to{\rightarrow}
\def\pa{\partial}

\def\lab{\label}

\newcommand{\polvec}{U}

\newcommand{\spin}{\ell}
\newcommand{\bbprop}{G}
\newcommand{\diffop}[5]{\mathbb{D}_{#1,#2}^{(#3)}(#4,#5)}
\newcommand{\bbnorm}[2]{a_{#1,#2}}

\newtoggle{editing}
\togglefalse{editing}

\iftoggle{editing}{%
	\usepackage[draft]{showlabels}

	\newcommand{\JSout}[1]{{\textbf{\textcolor{blue}{\sout{#1}}}}}
	\newcommand{\JSeqout}[1]{{\mathbf{\textcolor{blue}{\cancel{#1}}}}}

	\newcommand{\dan}[1]{{\textbf{\textcolor{purple}{[#1  --Dan]}}}}
	
	\newcommand{\ethan}[1]{{\textbf{\textcolor{red}{[#1  --Ethan]}}}}

}{%

	\newcommand{\JSout}[1]{}
	\newcommand{\JSeqout}[1]{}
	\newcommand{\dan}[1]{}
	\newcommand{\ethan}[1]{}
}

\begin{document}

\preprint{PUPT-2479}

\institution{Stanford}{Stanford Institute for Theoretical Physics, Stanford University, Stanford, \cr CA 94305, USA}
\institution{CTP}{Center for Theoretical Physics, Massachusetts Institute of Technology, \cr Cambridge, Massachusetts 02139, USA}
\institution{MITmath}{Department of Mathematics, Massachusetts Institute of Technology, \cr Cambridge, Massachusetts 02139, USA}
\institution{SLAC}{Theory Group, SLAC National Accelerator Laboratory, Menlo Park, \cr CA 94025, USA}
\institution{McGill}{Department of Physics, McGill University, Montr\'eal, QC H3A 2T8, \cr Canada}

\title{Spinning Geodesic Witten Diagrams}

\authors{Ethan Dyer,\worksat{\Stanford} Daniel Z. Freedman,\worksat{\Stanford,\CTP,\MITmath}  and James Sully\worksat{\SLAC,\McGill}}

\abstract{We present an expression for the four-point conformal blocks of symmetric traceless operators of arbitrary spin as an integral over a pair of geodesics in Anti-de Sitter space, generalizing the geodesic Witten diagram formalism of Hijano et al\cite{Hijano:2015zsa} to arbitrary spin. As an intermediate step in the derivation, we identify a convenient basis of bulk three-point interaction vertices which give rise to all possible boundary three point structures. We highlight a direct connection between the representation of the conformal block as geodesic Witten diagram and the shadow operator formalism.}
\date{}

\maketitle
\tableofcontents
\newpage
\section{Introduction}

The study of conformal field theory in general spacetime dimension $d$ has enjoyed a renaissance  in recent years  in part stimulated by the conformal bootstrap program (see \cite{Rattazzi:2008pe,Rychkov:2009ij,Caracciolo:2009bx,Poland:2010wg,Rattazzi:2010gj,Rattazzi:2010yc,Vichi:2011ux,Poland:2011ey,ElShowk:2012ht}  for example).  In this approach, pioneered by Mack and Salam \cite{Mack:1969rr}, Ferrara, Gatto, Grillo, Parisi \cite{Ferrara:1972xe,Ferrara:1972ay,Ferrara:1972uq,Ferrara:1973vz,Ferrara:1973eg}, and 
Polyakov \cite{Polyakov:1974gs}, and recently reviewed in \cite{Rychkov:2016iqz,Simmons-Duffin:2016gjk,Penedones:2016voo},  the consequences
of conformal and crossing symmetry are combined to derive important constraints on dynamics.  The AdS/CFT correspondence has also played a significant role for CFT's that possess a gravity dual in $d+1$ dimensions.  

Both approaches focus on the properties and computation of correlation functions of primary operators.  Conformal symmetry determines the spacetime dependence of two- and three-point correlators,  but does
not fix the OPE coefficients which are dynamical quantities. Four-point functions are more complicated.
They are in principle fixed by conformal symmetry along with knowledge of the OPE coefficients. In practice, however, one must compute the kinematic, four-point conformal blocks, to carry this out.

In the context of the AdS/CFT correspondence, the four-point function can be computed by summing bulk Witten diagrams with the four points $x_{i}$ located on the AdS boundary and internal vertices integrated over the entire AdS geometry.  In the series of papers, \cite{Hijano:2015rla,Hijano:2015zsa,Hijano:2015qja},  the authors asked and answered the question, ``What is the bulk dual of the conformal block?'' The elegant answer is that the block is precisely given by considering the tree-level exchange Witten diagram and restricting the bulk integrals to two geodesics, $\g_{12}$ and $\g_{34}$ which terminate at the boundary points $x_1,~x_2$ and $x_3,~x_4$, respectively.  This structure is called a geodesic Witten diagram.   This picture was established in \cite{Hijano:2015zsa} for external scalar operators and in \cite{Nishida:2016vds} for the case of a single external operator with spin exchanging a scalar.  In this paper the formalism is extended to internal and external operators of general integer spin $\ell$.  It should be noted that geodesic Witten diagrams give valid representations of conformal blocks whether or not the CFT has a holographic dual.

In our work we combine three main sources of information.
 We use  a modification of the general basis for three-point correlators of spinning operators obtained in \cite{Costa:2011mg} and the general treatment of  bulk propagators and their split representation in \cite{Costa:2014kfa}.  We develop the connection between the bulk geodesic Witten diagram presentation of the block and the shadow formalism used in \cite{SimmonsDuffin:2012uy}.  
 Our main technical innovation is the construction of a set of bulk vertices for which the geodesic Witten diagram
is easily evaluated.  For the general case of external operators with spin, this results in 1:1 correspondence between a set of preferred bulk structures and the three-point correlators. Ultimately, we use these novel three point bulk vertices to write an expression for the spinning conformal block as an integral over bulk geodesics. We cast our discussion in terms of the embedding space formalism \cite{Dirac:1936fq,Mack:1969rr,Boulware:1970ty,Ferrara:1973eg,Weinberg:2010fx} throughout. Other recent work on spinning conformal blocks can be found in \cite{Costa:2011dw, SimmonsDuffin:2012uy, Alday:2013cwa, Costa:2014rya, Penedones:2015aga, Rejon-Barrera:2015bpa, Iliesiu:2015qra, Iliesiu:2015akf, Echeverri:2015rwa, Costa:2016hju, Costa:2016xah, Echeverri:2016dun, Schomerus:2016epl, Fortin:2016dlj}.


The paper is organized as follows. Section~\ref{sec:rev} begins with a review of the conformal block decomposition and then summarizes the scalar geodesic Witten diagram of \cite{Hijano:2015zsa} in a notation conducive to generalization. The section concludes with a connection between the geodesic Witten expression for the block and the shadow formalism of \cite{SimmonsDuffin:2012uy,Ferrara:1972xe,Ferrara:1972ay,Ferrara:1972uq,Ferrara:1973vz}. In Section~\ref{sec:threepoint} we present the map between a preferred set of bulk three point vertices and boundary three point structures. Section~\ref{sec:spinning-blocks} is our main result, where we use the bulk three point vertices to construct spinning geodesic Witten diagrams for arbitrary external spins, and prove that this gives the spinning conformal block. In Section~\ref{sec:discussion} we conclude with discussion and future directions.
We have also have included three appendices. In Appendix~\ref{app:props} we establish conventions and collect identities for the various bulk-to-bulk and bulk-to-boundary propagators used throughout. In Appendix~\ref{app:monod} we elaborate on the monodromy projection used to isolate the contribution of an operator rather than its shadow to the conformal block. In the final appendix, Appendix~\ref{app:oddsends}, we collect a few derivations which did not fit in the main text.
\\
\*

\noindent While completing this work we became aware of a related paper by Alejandra Castro, Eva Llabr\'es, and Fernando Rejon-Barrera \cite{CastroEtAl}. The authors use different techniques to address similar questions.

\section{Review}\label{sec:rev}
It has been understood since the work of Dirac, that the $d$-dimensional conformal group, $SO(d+1,1)$, is quite powerful in fixing local correlation functions.\footnote{We will  be concerned exclusively with Euclidean theories. We use the notation of the embedding space formalism which is explained in Sec. \ref{sec:embedding-essentials}. }  Two-point functions of  (normalized) primary scalar operators are completely fixed, and their three-point functions are determined up to a constant:
\es{twoandthreepoint}{
\langle O(P_{1})O(P_{2})\rangle&=\frac{1}{P_{12}^{\Delta}}\\
\langle O_{1}(P_{1})O_{2}(P_{2})O_{3}(P_{3})\rangle&=\frac{C_{123}}{P_{12}^{\frac{\Delta_{1}+\Delta_{2}-\Delta_{3}}{2}}P_{13}^{\frac{\Delta_{1}+\Delta_{3}-\Delta_{2}}{2}}P_{23}^{\frac{\Delta_{2}+\Delta_{3}-\Delta_{1}}{2}}}\,,
}
where we have used the embedding space notation for the distance between two points: $P_{ij} = (y_i-y_j)^2$. 
In principle, the set of all three point coefficients, $C_{ijk}$ (and analogous couplings for spinning fields), completely fixes all local correlation functions in a conformal field theory. 

In practice, it is useful to decompose higher point functions in terms of structures  which are invariants of the conformal symmetry. For example, the four-point function of primary scalar operators may be written as\footnote{We will frequently use a hatted notation where, $\hat{f}=\left(\frac{P_{14}}{P_{13}}\right)^{\frac{\tau_{3}-\tau_{4}}{2}}\left(\frac{P_{24}}{P_{14}}\right)^{\frac{\tau_{1}-\tau_{2}}{2}}\frac{f}{{P_{12}^{\frac{\tau_{1}+\tau_{2}}{2}}P_{34}^{\frac{\tau_{3}+\tau_{4}}{2}}}}$, with $\tau=\Delta+\spin$.}
\es{fourpointfunction}{
\langle O_{1}(P_{1})O_{2}(P_{2})O_{3}(P_{3})O_{4}(P_{4})\rangle&=\left(\frac{P_{14}}{P_{13}}\right)^{\frac{\Delta_{3}-\Delta_{4}}{2}}\left(\frac{P_{24}}{P_{14}}\right)^{\frac{\Delta_{1}-\Delta_{2}}{2}}\frac{g_{\Delta_{i}}(u,v)}{{P_{12}^{\frac{\Delta_{1}+\Delta_{2}}{2}}P_{34}^{\frac{\Delta_{3}+\Delta_{4}}{2}}}}\\
&=\hat{g}_{\Delta_{i}}(P_i)\,,
}
where $g_{\Delta_{i}}(u,v)$  depends only on the conformal cross ratios, $u=\frac{P_{12}P_{34}}{P_{13}P_{24}}$, $v=\frac{P_{14}P_{23}}{P_{13}P_{24}}$.
The quantity $g(u,v)$ can be further decomposed into the sum of conformal blocks.\footnote{Explicit formulas for scalar blocks can be found in the work of Dolan and Osborn \cite{Dolan:2000ut,Dolan:2003hv}.}
\es{blockdecomp}{
g_{\Delta_{i}}(u,v)&=\sum_{\Delta,\spin}C_{12\Delta}C_{34\Delta}\bbprop_{\Delta_{i};\Delta,\spin}(u,v)\,.
}

Each conformal block contains the contribution of a given primary labelled by its dimension, $\D$, and spin, $\ell$, as well as its descendants, viz.
\es{blockdef}{
\hat{\bbprop}_{\Delta_{i};\Delta,\ell}(u,v)=\frac{1}{C_{12\Delta,\spin}C_{34\Delta,\spin}}\sum_{\alpha\in\mathcal{M}_{O_{\Delta,\spin}}}\langle O_{{1}}(P_{1})O_{{2}}(P_{2})|\alpha\rangle\langle\alpha |O_{{3}}(P_{3})O_{{4}}(P_{4})\rangle\,,
}
and is entirely determined by conformal symmetry, while the dynamical information is contained in the three-point OPE  coefficients.
In writing \eqref{blockdef} we have chosen an orthonormal basis for the states $\alpha$  in the conformal family of $O_{\Delta}$. 
The hatted object, $\hat \bbprop_{\Delta,\spin}(P_i)$, is typically referred to as the conformal partial wave.  

For operators with spin, the essential structure is the same, although the details are more elaborate.\footnote{In this paper, we will discuss symmetric traceless tensor operators. 
 However we believe the results will extend to conformal blocks of all representations.} 
The two-point function of an operator with spin is again unique up to normalization, and is given by,
\es{twopoint}{
\langle O(P_{1},\polvec_{1})O(P_{2},\polvec_{2})\rangle&=\frac{H_{12}^{\spin}}{P_{12}^{\tau}}\,.
}
Here we have introduced the notation, $\tau=\Delta+\spin$, and contracted the indices of $O$ with a polarization vector $O(P,\polvec)\equiv \polvec_{A_{1}}\polvec_{A_{2}}\ldots \polvec_{A_{\spin}}O^{A_{1}A_{2}\ldots A_{\spin}}(P)$.  The scalar, $H_{ij}$, depends on the positions, $P_{i}$, and polarization vectors, $\polvec_{i}$.
\es{Hdef}{
H_{ij}&=-\tr\left(C_{i}C_{j}\right)=-2[(U_i\cdot U_j)(P_i\cdot P_j)- (U_i\cdot  P_j)(U_j\cdot P_i)]\,.
}
This is written in terms of the useful intermediate structure, $C_{iAB}=\polvec_{iA}P_{iB}-\polvec_{iB}P_{iA}$. We elaborate on this notation and the analogue for bulk fields in Section \ref{sec:embedding-essentials}.

For three-point functions of operators with fixed, non-zero spin, we no longer have a single conformal invariant as in \reef{twoandthreepoint}.  Instead, as explained in \cite{Costa:2011mg} and reviewed in Sec. \ref{sec:threepoint} below, there is a finite-dimensional space of three-point functions spanned by a set of three-point structures.
Denoting the independent structures by $\mathcal{V}^{I}(P_{1},\polvec_{1};P_{2},\polvec_{2};P_{3},\polvec_{3})$, we can write the general expression for a three point function of spinning operators as the sum
\es{tpt}{
\langle O_{{1}}(P_{1},\polvec_{1})O_{{2}}(P_{2},\polvec_{2})O_{{3}}(P_{3},\polvec_{3})\rangle&=C_{123;I}\frac{\mathcal{V}^{I}(P_{1},\polvec_{1};P_{2},\polvec_{2};P_{3},\polvec_{3})}{P_{12}^{\frac{\tau_{1}+\tau_{2}-\tau_{3}}{2}}P_{13}^{\frac{\tau_{1}+\tau_{3}-\tau_{2}}{2}}P_{23}^{\frac{\tau_{2}+\tau_{3}-\tau_{1}}{2}}}\,.
}
As in the scalar case, we can decompose the four-point function of spinning operators in terms of the three-point coefficients, which specify the dynamical data, and the spinning conformal blocks, which encapsulate the kinematics.
\es{spin4pt}{
\langle O_{{1}}(P_{1},\polvec_{1})O_{{2}}(P_{2},\polvec_{2})O_{{3}}(P_{3},\polvec_{3})O_{{4}}(P_{4},\polvec_{4})\rangle&=\left(\frac{P_{14}}{P_{13}}\right)^{\frac{\tau_{3}-\tau_{4}}{2}}\left(\frac{P_{24}}{P_{14}}\right)^{\frac{\tau_{1}-\tau_{2}}{2}}\frac{g_{\Delta_{i},\spin_{i}}(P_{i},U_{i})}{{P_{12}^{\frac{\tau_{1}+\tau_{2}}{2}}P_{34}^{\frac{\tau_{3}+\tau_{4}}{2}}}}\,.
}
The function $g_{\Delta_{i},\spin_{i}}$ can be decomposed in terms of spinning conformal blocks.
\es{spinconfdecomp}{
g_{\Delta_{i},\spin_{i}}(P_{i},U_{i})&=\sum_{\Delta,\spin}C_{12\Delta,\spin;I}C_{34\Delta,\spin;J}\bbprop_{\Delta_{i},\spin_{i};\Delta,\spin}^{IJ}(P_{i},\polvec_{i})\,.
}
In this paper, we present an expression for $\bbprop_{\Delta_{i},\ell_{i};\Delta,\spin}^{IJ}$ in terms of geodesic Witten diagrams.
The technology of geodesic Witten diagrams was first developed for scalar four-point functions in \cite{Hijano:2015zsa}, and we review this construction below.

\subsection{Geodesic Witten Diagrams}

\begin{figure}
	\centering
	\includegraphics[width=0.5\linewidth]{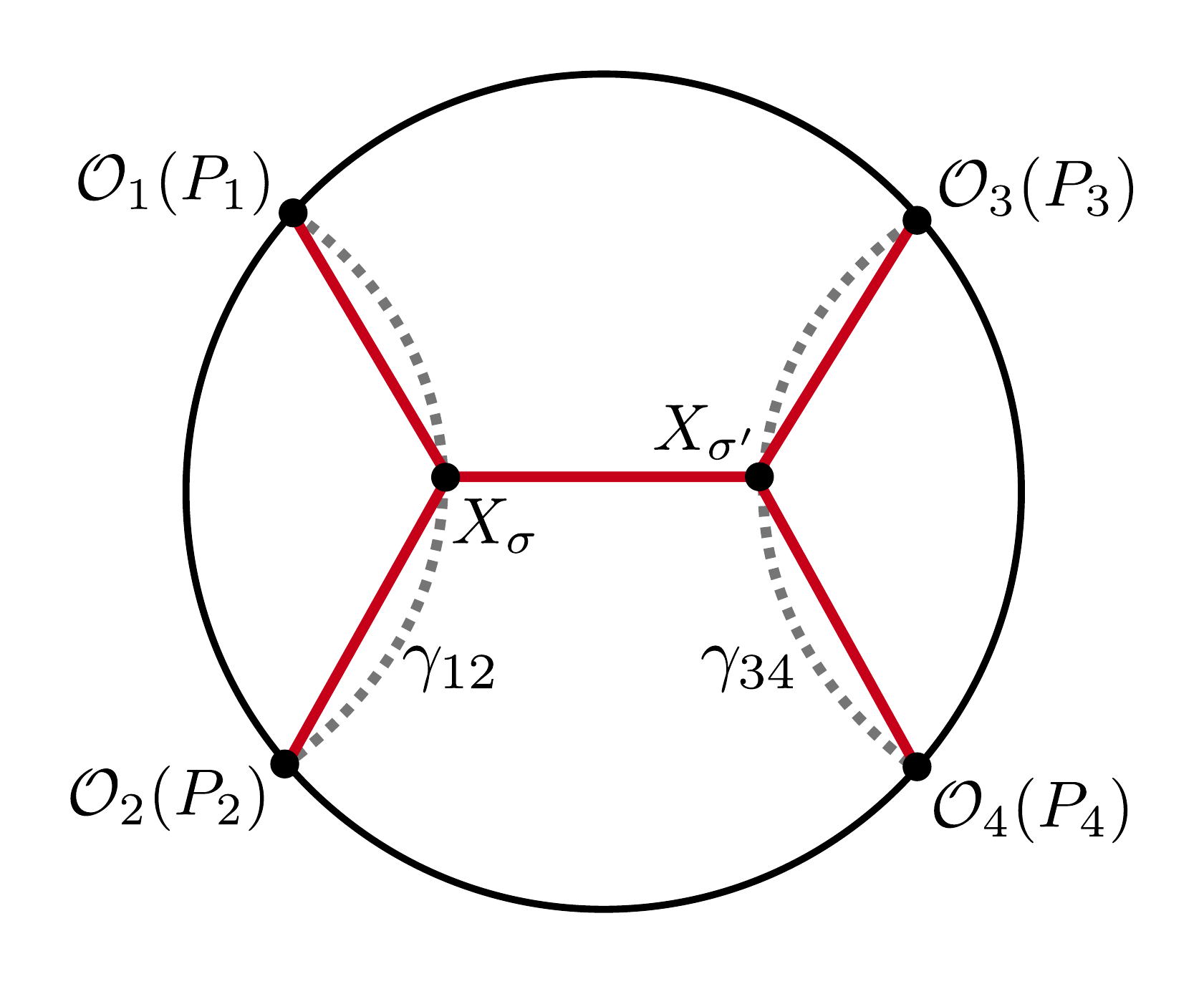}
	\caption{A geodesic Witten diagram (GWD) for a scalar conformal block. The GWD consists of a regular exchange Witten diagram, where the interaction vertices are restricted to lie on the geodesics $\gamma_{12}$,$\gamma_{34}$ connecting the boundary operators.}
	\label{fig:gwdsc}
\end{figure}

The geodesic Witten diagram (GWD) provides a dual, bulk description of the conformal block. The simplest example is the GWD corresponding to the conformal block for four scalars exchanging an intermediate scalar operator. This is represented in terms of an exchange Witten diagram where the vertices are integrated over geodesics connecting the boundary points, rather than over the entire bulk, see Figure~\ref{fig:gwdsc}.

Quantitatively, the relationship is,
\es{eq:simpgeodwit}{
\hat{\bbprop}_{\Delta_{i},\Delta}(u,v)&=\int_{\gamma_{12}} d\sigma \int_{\gamma_{13}}d\sigma^{\prime} K_{\Delta_{1}}(P_{1};X_{\sigma})K_{\Delta_{2}}(P_{2};X_{\sigma})\bbprop_{\Delta}(X_{\sigma};X_{\sigma^{\prime}})K_{\Delta_{3}}(P_{3};X_{\sigma^{\prime}})K_{\Delta_{4}}(P_{4};X_{\sigma^{\prime}})\,.
}
Here, the paths $\gamma_{12}$ and $\gamma_{34}$ are geodesics between the pairs of boundary points, parametrized by the proper lengths, $\sigma$ and $\sigma^{\prime}$. $K_{\Delta}(P;X)$ is a bulk-to-boundary propagator, defined in Appendix~\ref{app:props}, and in a suggestive abuse of notation, $\hat\bbprop$ on the LHS of the equation represents the partial-wave, and on the RHS represents the bulk-to-bulk propagator.

In \cite{Hijano:2015zsa}, the authors show that the GWD gives the conformal block by demonstrating that it satisfies the correct Casimir equation,
\es{caseq}{
\frac{1}{2}(\mathcal{L}_{AB}^{1}+\mathcal{L}_{AB}^{2})^{2}\hat{\bbprop}_{\Delta_{i},\Delta}(u,v)&=-C_{2}(\Delta,0)\hat{\bbprop}_{\Delta_{i},\Delta}(u,v)\,,
}
and that it gives the correct behavior in the limit $u \to 0$;
\es{shortdist}{
\bbprop_{\Delta_{i},\Delta}(u,v)\sim u^{\D/2}\,.
}
Here, $C_{2}(\Delta,\spin)=\D(\D-d) +\spin(\spin+d-2)$.

We will demonstrate that the spinning GWD also has the correct short-distance behavior, and satisfies the Casimir equation. We also introduce an alternative perspective on why the geodesic Witten diagram gives the conformal block. This involves a natural relation between the expression for the conformal block as a geodesic Witten diagram, and the shadow formalism introduced in \cite{Ferrara:1972xe,Ferrara:1972ay,Ferrara:1972uq,Ferrara:1973vz}, and used in \cite{SimmonsDuffin:2012uy}, which we discuss below.

\subsection{Connection to Shadow Formalism}  \label{sec: shadow1}

In  \cite{SimmonsDuffin:2012uy}, conformal blocks were expressed as a projector acting on a product of three-point functions integrated over the boundary of AdS$_{d+1}$. For the case of scalar conformal blocks, this is done by first introducing an object, $\Omega$ with simple scaling properties, \es{eq:shadowint}{
\hat{\Omega}_{\Delta_{i};\Delta}(u,v)&=\int d^{d}P\langle O_{1}(P_{1})O_{2}(P_{2})O(P)\rangle\langle\tilde{O}(P)O_{3}(P_{3})O_{4}(P_{4})\rangle\,.
}
Here, $O$ is a scalar operator of dimension $\Delta$, and $\tilde{O}$ is its shadow with dimension $d-\Delta$. This integral can be evaluated in terms of the conformal blocks, see Appendix~\ref{app:confint}
\es{shadowIntRes}{
\Omega_{\Delta_{i};\Delta}(u,v)&\propto\left(\bbprop_{\Delta_{i};\Delta}(u,v)-\bbprop_{\Delta_{i};d-\Delta}(u,v)\right)\,.
}
The expression in parenthesis is the sum of the conformal block for the operator, $O$, and the shadow.
The direct block and shadow both satisfy the same Casimir equation, but can be distinguished by their behavior in the limit $u\rightarrow0$.

In order to pick out the block rather than the shadow, one can project onto the correct singular behavior. This is done by introducing a \textit{monodromy projection} (selecting the term with the correct monodromy as $u$ circles zero). Said  more concretely, this projection picks the terms of the form $u^{\Delta/2}$ and sets to zero the terms of the form $u^{(d-\Delta)/2}$ in the expansion around $u=0$.
\es{projection}{
\bbprop_{\Delta_{i};\Delta}(u,v)&=\mathcal{P}_{\Delta}\Omega_{\Delta_{i};\Delta}(u,v)\,.
}

\begin{figure}
	\centering
	\includegraphics[width=0.9\linewidth]{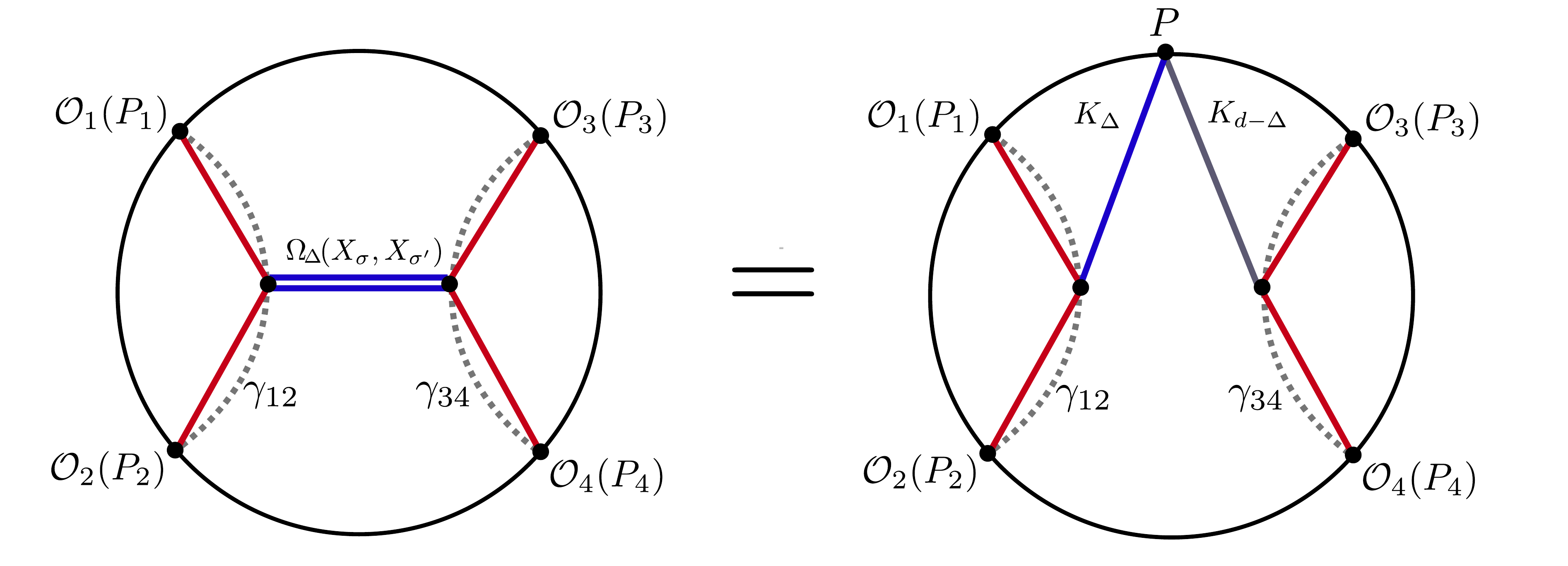}
	\caption{The geodesic Witten diagram (GWD) for the harmonic function. On the LHS, the bulk propagator in the regular GWD has been replaced by a bulk harmonic function, represented by the double line. On the RHS, an equivalent representation for the bulk harmonic function is given in the `split representation' where bulk-to-boundary operators for the bulk field and its shadow are integrated over the boundary, as in Eq. \eqref{simpsplitrep}.}
	\label{fig:gwd-harmonic}
\end{figure}

To make connection with the GWD formalism, we  describe the bulk diagram which reproduces $\Omega_{\Delta_{i};\Delta}(u,v)$, depicted in Fig. \ref{fig:gwd-harmonic} and given by the integral 
\es{eq:simpgeodwit-2}{
\hat{\Omega}_{\Delta_{i},\Delta}(u,v)&=\int_{\gamma_{12}} d\sigma \int_{\gamma_{34}}d\sigma^{\prime} K_{\Delta_{1}}(P_{1};X_{\sigma})K_{\Delta_{2}}(P_{2};X_{\sigma})\Omega_{\Delta}(X_{\sigma};X_{\sigma^{\prime}})K_{\Delta_{3}}(P_{3};X_{\sigma^{\prime}})K_{\Delta_{4}}(P_{4};X_{\sigma^{\prime}})\,.
}
This is similar to \reef{eq:shadowint}, except that  the bulk-to-bulk propagator is replaced by 
 the \emph{bulk} harmonic function, $\Omega_{\Delta}(X;X^{\prime})$, 
 defined as the difference between the bulk propagators of the fields dual to  $O_\D$ and its shadow $O_{d-\D}$:
\es{decompofbb}{
\Omega_{\Delta}(X;X^{\prime})&=\bbprop_{\Delta}(X;X^{\prime})-\bbprop_{d-\Delta}(X;X^{\prime})\,.
}
%
%
%
To show that this reproduces the boundary shadow integral, \eqref{eq:shadowint}, we use the split representation of the bulk harmonic function whose integrand contains the product of bulk-to-boundary propagators for the operator of interest and its shadow,
\es{simpsplitrep}{
\Omega_{\Delta}(X;X^{\prime})=(d-2\D)\int d^{d}P K_{\Delta}(P;X)K_{d-\Delta}(P;X^{\prime})\,.
}
With this representation, the integrals over the bulk geodesics can be easily performed. The product of three bulk-to-boundary propagators integrated over a bulk geodesic connecting two boundary points is proportional to a boundary three-point function.
\es{threepointint}{
\int_{\gamma_{12}}d\sigma K_{\Delta_{1}}(P_{1};X_{\sigma})K_{\Delta_{2}}(P_{2};X_{\sigma})K_{\Delta_{3}}(P_{3},X_{\sigma})\propto \langle O_{2}(P_{2})O_{1}(P_{1})O_{3}(P_{3})\rangle\,.
}
Using this relation, and plugging \eqref{simpsplitrep} into \eqref{eq:simpgeodwit-2}, the first integral, over $\gamma_{12}$, produces the three point function of $O_{1}$, $O_{2}$, and $O$, while the other, over $\gamma_{34}$, produces the three point function of the shadow, $\tilde{O}$ with $O_3,~O_4$. The remaining integral over the boundary is exactly the integral in \eqref{eq:shadowint}.
%
%

We are now in a position to compare the shadow and GWD presentation of the block. In the shadow prescription we apply a projector on the boundary selecting the pieces of $\hat{\Omega}_{\Delta_{i},\Delta}(u,v)$ with the correct behavior around the $u\rightarrow 0$ limit. 

\begin{figure}
	\centering
	\includegraphics[width=0.7\linewidth]{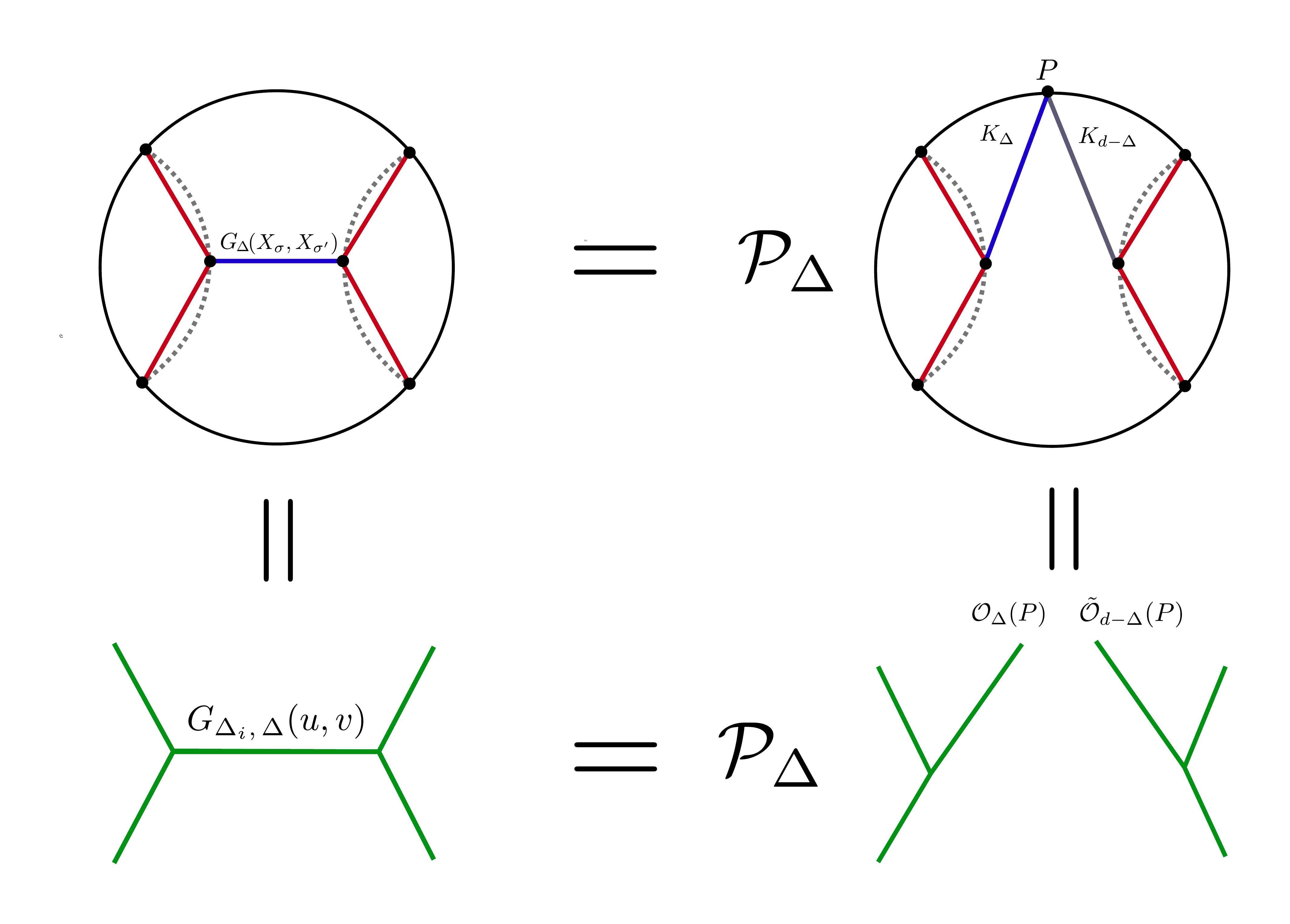}
	\caption{The geodesic Witten diagram is recovered from acting on the bulk harmonic with a suitable projector. This is equivalent to acting with a projector on the shadow-operator representation of the harmonic function, and demonstrates the connection between the conformal block and the GWD.}
	\label{fig:bulk-proj-intro}
\end{figure}

In the GWD presentation, we use the bulk-to-bulk propagator, instead of the bulk harmonic function. But the bulk-to-bulk propagator can also be cast in terms of projectors, in this case acting on the bulk harmonic function. 
This projector picks out the direct propagator, rather than the shadow, in \eqref{decompofbb}, by selecting the correct behavior around the $-X\cdot X^{\prime}\rightarrow\infty$ limit. In this limit, the bulk-to-bulk propagator behaves as,
\es{b2blim}{
\bbprop_{\Delta}(X;X^{\prime})&\sim \frac{1}{(-X\cdot X^{\prime})^{\Delta}}\,,
}
so we can keep the direct propagator, and discard the shadow propagator in \eqref{decompofbb} by projecting onto terms of the form $(-X\cdot X^{\prime})^{-\Delta-2n}$ with integer $n$.
\es{bulkprojection}{
\bbprop_{\Delta}(X,X^{\prime})&=\mathcal{P}_{\Delta}\Omega_{\Delta}(X,X^{\prime})\,.
}
This looks very similar to the boundary projection \eqref{projection}, but now acting on bulk objects. The geodesic integrals convert the appropriate branch structure in $X\cdot X^{\prime}$ to the corresponding structure in $u$, and thus the bulk and boundary projectors map as
\es{simpgeodwitproj}{
\mathcal{P}_{\Delta}\hat{\Omega}_{\Delta_{i},\Delta}(u,v)&=\int_{\gamma_{12}} d\sigma \int_{\gamma_{34}}d\sigma^{\prime} K_{\Delta_{1}}(P_{1};X_{\sigma})K_{\Delta_{2}}(P_{2};X_{\sigma})\mathcal{P}_{\Delta}\Omega_{\Delta}(X_{\sigma};X_{\sigma^{\prime}})K_{\Delta_{3}}(P_{3};X_{\sigma^{\prime}})K_{\Delta_{4}}(P_{4};X_{\sigma^{\prime}})\\
&=\int_{\gamma_{12}} d\sigma \int_{\gamma_{34}}d\sigma^{\prime} K_{\Delta_{1}}(P_{1};X_{\sigma})K_{\Delta_{2}}(P_{2};X_{\sigma})\bbprop_{\Delta}(X_{\sigma};X_{\sigma^{\prime}})K_{\Delta_{3}}(P_{3};X_{\sigma^{\prime}})K_{\Delta_{4}}(P_{4};X_{\sigma^{\prime}})\,,
}
where the second line is precisely the geodesic Witten diagram (see Figure \ref{fig:bulk-proj-intro}). This shows how the split representation of the bulk-to-bulk propagator, \eqref{simpsplitrep}, can be used to demonstrate a simple equivalence between the GWD and boundary shadow integral presentation of the conformal block. Though the details become somewhat more elaborate, the same basic relation holds in the spinning case as well.

\subsection{Embedding Space Essentials}\label{sec:embedding-essentials}

The notation of the embedding space formalism has already been used above, and it is essential in the rest of this paper.  Therefore we attempt to give a minimal set of rules for calculations in embedding space. Our discussion below is a distillation of more detailed reviews in  \cite{Costa:2011mg,SimmonsDuffin:2012uy,Costa:2014kfa}.

One can extend the $d$-dimensional physical configuration space of a CFT to the $d+2$-dimensional \emph{embedding space} of  signature $(d+1,1)$ so that the $SO(d+1,1)$ group acts linearly on embedding space vectors $X^A$.   One uses the signature $+\ldots +-$ Cartesian metric in light cone coordinates, so that
the scalar product of two vectors is
\be\lab{scprod}
X\cdot X' = -\tfrac12(X^+X'^-+X^-X'^+) + X^i X'^i\,.
\ee

Points in the physical space of the CFT$_d$ correspond to null rays $\l P^A\big |_{y}$ with $P^A\big |_{y}= (1,y^2, y^i)$, $\l~ \text real.$
Embedding space is also useful for for describing AdS, as Euclidean AdS$_{d+1}$ (with scale $L=1$) is naturally viewed as the hyperboloid $X^2 = -1$ in the interior of the forward cone.  The mapping from the embedding space to the Poincar\'e patch of AdS is realized by vectors of the form $X^A = (1, x_0^2 +x^2, x^i)/x_0$ while the ray through $P^A$ can be identified  as the AdS boundary.   Scalar products  in embedding space are related to  familiar quantities in the Poincar\'e patch and its boundary:
\be
P_{12} \equiv -2 P_1\cdot P_2 = (y_1-y_2)^2 \qquad\qquad  -2X\cdot P= (x_0^2 +(x-y)^2))/x_0 .   
\ee

\paragraph{CFT fields with spin} Embedding space methods are especially convenient for fields and operators with spin.  A boundary conformal field with scale dimension dimension $\D$ and spin $\spin$ is described by a:
\begin{itemize}
	\item \textbf{symmetric, traceless rank $\spin$ tensor} $F_{A_1,\ldots,A_\spin}(P)$.
	\item  This tensor is \textbf{homogeneous of degree $-\D$}, i.e.  $F_{A_1,\ldots,A_\spin}(\l P) = \l^{-\D} F_{A_1,\ldots,A_\spin}(P)$
	\item  and \textbf{transverse}, $P^{A_1} F_{A_1,\ldots,A_\spin}=0$.
\end{itemize}
The tensor is defined on the cone, and its projection to the physical space is  
\be\lab{projten}
f_{i_1,\ldots, i_\spin}(y) = \frac{\pa P^{A_1}}{\pa y^{i_1}} \ldots \frac{\pa P^{A_1}}{\pa y^{i_\spin}}F_{A_1,\ldots,A_\spin}\left(P\big |_{y}\right),
\qquad\quad \frac{\pa P^{A}}{\pa y^i} = (0,2y_i,\d^A_i).
\ee
Any tensor $F_{A_1,\ldots,A_\spin}$ proportional to $P_A$ projects to zero and may be called a "pure gauge."   
The tensor  $f_{i_1,\ldots, i_\spin}$ is traceless if its parent $F_{A_1,\ldots,A_\spin}$ is traceless and transverse.


\paragraph{AdS fields with spin} Likewise, symmetric traceless fields in the Poincar\'e patch of AdS$_{d+1}$ can be described as 
\begin{itemize}
	\item \textbf{symmetric, traceless tensors} $H_{A_1,\dots A_\spin}(X)$ in embedding space,
	\item  and are \textbf{transverse}, $X^{A_1} H_{A_1,\dots A_\spin}(X) =0$. 
\end{itemize}
The tensors are defined on the hyperboloid $X^2=-1$, and any extension away from this AdS hypersurface is unphysical. Hence tensors that are not explicitly transverse must be projected to the tangent plane using the 
induced metric $\Pi_{AB} = \eta_{AB} +X_AX_B$ with one index raised to form the projector $\Pi_A{}^B = \Pi_{AC}\eta^{CB}$. The projection to the Poincar\'e patch is given by
\bea\lab{bulkproj}
h_{\m_1,\dots \m_\spin} &=& \frac{\pa X^{A_i}}{\pa x^{\m_i}}\ldots  \frac{\pa X^{A_\spin}}{\pa x^{\m_\spin}} H_{A_1,\dots A_\spin}\,.
\eea


\paragraph{Embedding space polynomials} Tensor indices are difficult to deal with.  Fortunately this problem can be eased by packaging symmetric traceless tensor fields in both physical and embedding space in polynomials using polarization vectors.  
This formalism leads to economies of notation and calculation.
For boundary and bulk fields we write:
\es{eq:poly}{
F(P,\polvec)&=F_{A_1,\ldots,A_\spin}(P)\,\polvec^{A_1}\ldots \polvec^{A_\spin}\\
 H(X,W) &= \,H_{A_1,\dots A_\spin}(X)\,W^{A_1}\ldots W^{A_\spin}\,.
}
In embedding space it is convenient to encode transversality and traceless-ness by taking the boundary polarization vectors, $\polvec^A$ ,to satisfy $\polvec\cdot \polvec= P\cdot \polvec=0$, and the bulk polarization vectors, $W^A$, to satisfy $W \cdot W = X \cdot W=0$.

\paragraph{Freeing indices} To recover a symmetric, traceless boundary tensor from an arbitrary embedding space polynomial, and free indices, we use the operator
\es{eq:free-bound-index}{
D_A =  \left( \frac{d-2}{2}  + \polvec \cdot \frac{\partial}{\partial \polvec}\right)\frac{\partial}{\partial \polvec^A}- \frac{1}{2} \polvec_A \frac{\partial^2}{\partial \polvec\cdot \partial \polvec} \, .
}
This has the effect of simultaneously freeing the index and inserting a projector. To contract two tensors, say $F(P,\polvec)$ and $G(P,\polvec)$, we simply evaluate 
\es{eq:contract}{
F(P,D)G(P,\polvec) \; .
} 
Similarly, we can free bulk indices using an operator $K_A$. 
This operator, however, is more more complicated, and its specific form will not be needed in this paper (it can be found in (12) of \cite{Costa:2014kfa}). 
We emphasize only that $K_A$ is defined so that
\es{eq:k-op-job}{
\frac{1}{\ell!\left(\tfrac{d-1}{2}\right)!}K_{A_1} \ldots K_{A_\ell} W^{B_1} \ldots W^{B_\ell} = \Pi_{\lbrace A_1}^{B_1} \ldots \Pi_{ A_\ell \rbrace}^{B_\ell} \, ,
}
where again $\Pi_{AB} = \eta_{AB} + X_A X_B$ is the projector onto the AdS submanifold,  and $\{\dots\}$ includes both symmetrization and subtraction of traces.

\paragraph{Covariant derivative} The AdS covariant derivative acting on a tensor with free indices can be formed by projecting the partial derivative onto the transverse subspace using $\Pi_{AB}$. 
\es{covdfreeind}{
\nabla_B H_{A_1,\dots A_\spin}(X)=\Pi_B{}^C \Pi_{A_1}{}^C\dots \Pi_{A_\spin}{}^{C_\spin}\pa_C H_{C_1,\dots C_\spin}(X),
}
When acting on tensors that have all indices contracted with polarization vectors, this takes the simple form,
\es{eq:covariant-d}{
\nabla_A = \frac{\partial}{\partial X^A} + X_A \left(X \cdot \frac{\partial}{\partial X}\right) + W_A \left(X \cdot \frac{\partial}{\partial W}\right) \, .
}
The covariant derivative simplifies further when contracted with a polarization vector so that $W \cdot \nabla = W \cdot \partial_X$. The extra terms in the covariant derivative, which insert the projector, are then recovered when the indices are later freed by acting with $K_A$. 

\section{Three-Point Functions}\label{sec:threepoint}

In this section we discuss the possible tensor structures that can appear in the three-point functions of operators with spin. 
We then define a set of cubic bulk vertices that generate the same three-point functions through geodesic Witten diagram computations.

\subsection{CFT Three-Point Structures}

First, we review the basis of three-point tensor structures introduced in \cite{Costa:2011mg}, following the treatment of this paper closely. We then define a modified basis that is more convenient for our purposes.

Let us begin with a few illustrative examples. As the simplest spinning case, consider the three-point function of two scalars and a spin-$1$ operator. 
The form of the three-point function can be fixed analogously to the scalar case to take the form
\begin{equation}
	\langle O_{{1}}(P_{1})O_{{2}}(P_{2})O_{{3}}(P_{3})_{A}\rangle =\frac{\mathcal{V}(P_i)_{A}}{P_{12}^{\frac{\Delta_{1}+\Delta_{2}-\Delta_{3}}{2}}P_{13}^{\frac{\Delta_{1}+\Delta_{3}-\Delta_{2}}{2}}P_{23}^{\frac{\Delta_{2}+\Delta_{3}-\Delta_{1}}{2}}}\,.
\end{equation}
where the numerator accounts for the spin and must be a vector invariant under rescalings. 
What can appear in the numerator? Ignoring terms proportional to $P_3$, which vanish when pulled back to the physical space, the most general scale-invariant vector takes the form,
\begin{equation}
\mathcal{V}(P_i)_{A} = \alpha P_{1 \,A} \sqrt{\frac{P_{23}}{P_{13}P_{12}}} + \beta P_{2 \, A} \sqrt{\frac{P_{13}}{P_{23}P_{12}}} \, .
\end{equation}
Enforcing transversality, $P_3 \cdot O_{{3}}(P_{3}) = 0$, gives
\begin{equation}
\mathcal{V}(P_i)_{A} \propto  \left( P_{1 \, A} \sqrt{\frac{P_{23}}{P_{13}P_{12}}} - P_{2 \, A} \sqrt{\frac{P_{13}}{P_{23}P_{12}}} \right) \, .
\end{equation}
The unique scalar-scalar-vector three-point function then takes the form,
\begin{equation}\label{eq:spin-example-ssv}
	\langle O_{{1}}(P_{1})O_{{2}}(P_{2})O_{{3}}(P_{3})_{A}\rangle =\frac{C_{123} \, {\mathcal{V}}(P_i)_{A}}{P_{12}^{\frac{\tau_{1}+\tau_{2}-\tau_{3}}{2}}P_{13}^{\frac{\tau_{1}+\tau_{3}-\tau_{2}}{2}}P_{23}^{\frac{\tau_{2}+\tau_{3}-\tau_{1}}{2}}}\; ,
\end{equation}
where we recall $\tau = \Delta + \spin$ and we have set
\begin{equation}
\mathcal{V}(P_i)_{A} = \frac{P_{1 \, A} P_{23}- P_{2 \, A}P_{13}}{P_{12}} \, .
\end{equation}
The next step is obtain the corresponding embedding-space polynomial.  In this simple case we just  contract with the polarization vector $\polvec_3$:
\begin{equation}\label{Vssv}
\mathcal{V}(P_i,\polvec_3) = \frac{(\polvec_3 \cdot P_1) P_{23}- (\polvec_3 \cdot P_2)P_{13}}{P_{12}} \; .
\end{equation}
This particular spin structure is an important building block for the general basis, so we label it as follows:
\begin{equation}
V_{j,i k} \equiv \frac{P_i \cdot C_j \cdot P_k}{P_i \cdot P_k} =  \frac{ (\polvec_j \cdot P_i) (P_j \cdot P_k) - (\polvec_j \cdot P_k)(P_j \cdot P_i)}{(P_i \cdot P_k)} \, .
\end{equation}
Notice that $V_{j,ik}$ is transverse; it vanishes when $U_j \to P_j$.

This result can be generalized to the slightly more complicated example of the scalar-scalar-spin $\spin$ correlator. We immediately find
\begin{equation}\label{eq:spin-example-ssj}
\langle O_{{1}}(P_{1})O_{{2}}(P_{2})O^{(\spin)}_{{3}}(P_{3},\polvec_3)\rangle =\frac{C_{123} \, (V_{3,12})^\spin}{P_{12}^{\frac{\tau_{1}+\tau_{2}-\tau_{3}}{2}}P_{13}^{\frac{\tau_{1}+\tau_{3}-\tau_{2}}{2}}P_{23}^{\frac{\tau_{2}+\tau_{3}-\tau_{1}}{2}}}\; ,
\end{equation}
where the numerator now follows by imposing transversality in all $\spin$ indices. 

These cases are straightforward since there is unique tensor structure that can appear. What happens when the three operators have arbitrary spin? This was worked out in \cite{Costa:2011mg} and will be summarized next.

\subsubsection{The General Three-Point Function}

A three-point function of three operators of arbitrary spin takes the schematic form
\begin{equation}
\label{eq:spin-scheme-jjj}
\langle O^{(\spin_1)}_{{1}}(P_{1},\polvec_1)O^{(\spin_2)}_{{2}}(P_{2},\polvec_2)O^{(\spin_3)}_{{3}}(P_{3},\polvec_3)\rangle =\frac{\mathcal{V}^{(\spin_1,\spin_2,\spin_3)}(P_i,\polvec_i)}{P_{12}^{\frac{\tau_{1}+\tau_{2}-\tau_{3}}{2}}P_{13}^{\frac{\tau_{1}+\tau_{3}-\tau_{2}}{2}}P_{23}^{\frac{\tau_{2}+\tau_{3}-\tau_{1}}{2}}}\; ,
\end{equation}
The polynomial $\mathcal{V}^{(\spin_1,\spin_2,\spin_3)}(P_i,\polvec_i)$ must have the following properties:
\begin{itemize} 
\item{ To describe three operators with spin $\ell_i$, the polynomial must be a homogeneous polynomial of degree $(\spin_1,\spin_2,\spin_3)$ in $(\polvec_1,\polvec_2,\polvec_3)$;}
\item{ Correct scaling under $P_i\to\lambda P_i$ requires it to be homogeneous of degree  $\ell_i$  in each $P_i$;}
\item{And, it must also be transverse in each $P_i$.}
\end{itemize} 
A basis of for such polynomials can be built from monomials of the scalar structures,
\begin{eqnarray}
V_i = V_{i,jk}  &=&  \frac{P_j\cdot C_i \cdot P_k}{P_j\cdot P_k}  =  \frac{ (\polvec_i \cdot P_j) (P_i \cdot P_k) - (\polvec_i \cdot P_k)(P_i \cdot P_j)}{(P_j \cdot P_k)} \nonumber\\
H_{ij}  &=& - \tr(C_i C_j)  =  -2\left[(\polvec_i \cdot \polvec_j)(P_i \cdot P_j) - (\polvec_i \cdot P_j)(P_i \cdot \polvec_j)\right] \; ,
\end{eqnarray}
where
\begin{equation}
	C_{i\, AB} = \polvec_{i \, A} P_{i \, B} - P_{i \, A} \polvec_{i \, B} \, .
\end{equation}
Then we have a linear basis of polynomials for spins $(\spin_1,\spin_2,\spin_3)$ given by
\begin{equation}\label{oldbasis}
\mathcal{V}_I(P_i,\polvec_i) = \prod_i V_i^{m_i} \prod_{i<j} H^{n_{ij}}_{ij} \, ,
\end{equation}
where the integer exponents $m_i,~n_{ij}$ obey
\begin{equation}
m_i + \sum_{j \neq i } n_{ij} = \spin_i \, .
\end{equation}
Letting $I$ denote the set of ${\lbrace m_i,n_{jk}\rbrace}$ obeying this constraint, the three-point function takes the form
\begin{equation}\label{eq:spin-final-jjj}
\langle O^{(\spin_1)}_{{1}}(P_{1},\polvec_1)O^{(\spin_2)}_{{2}}(P_{2},\polvec_2)O^{(\spin_3)}_{{3}}(P_{3},\polvec_3)\rangle =\frac{\sum_{I} C_{123}^I \mathcal{V}_I (P_i,\polvec_i)}{P_{12}^{\frac{\tau_{1}+\tau_{2}-\tau_{3}}{2}}P_{13}^{\frac{\tau_{1}+\tau_{3}-\tau_{2}}{2}}P_{23}^{\frac{\tau_{2}+\tau_{3}-\tau_{1}}{2}}}\; ,
\end{equation}
and so is determined by a set of OPE coefficients, $C_{123}^I$, one for each possible tensor structure.
A more complete discussion of this basis of  boundary structures, including a count of independent structures, is given in Sec. 4 of \cite{Costa:2011mg}.

\subsubsection{An Alternate Tensor Basis}


We will employ an alternate basis, given by a linear combination of the structures of \reef{oldbasis}.
In this basis, the correspondence between boundary structures and bulk geodesic diagrams is much simpler.
The new basis is
\es{struc}{
	\tilde{\mathcal{V}}_I&=\sum_{a=0}^{n_{13}}\sum_{b=0}^{n_{23}}f(\kappa,\kappa^{\prime};a,b)V_{1,23}^{m_{1}+a}V_{2,13}^{m_{2}+b}V_{3,12}^{m_{3}+a+b}H_{12}^{n_{12}}H_{13}^{n_{13}-a}H_{23}^{n_{23}-b}\,,
}
with,
\es{ref}{
	f(\kappa,\kappa^{\prime};a,b)&=(-1)^{b}\frac{B(\kappa+a,\kappa^{\prime}+b)}{4^{n_{12}+n_{13}+n_{23}-2a-2b}}\left(\begin{array}{c}n_{13}\\a\end{array}\right)\left(\begin{array}{c}n_{23}\\b\end{array}\right)\\
	\kappa&=m_2-n_{13}+\frac{\tau_{1}+\tau_{3}-\tau_{2}}{2}\\
	\kappa^{\prime}&=m_{1}-n_{23}+\frac{\tau_{2}+\tau_{3}-\tau_{1}}{2}\,.
}
Here,
\es{Beta}{
	B(a,b)\,=\,\frac{\Gamma (a) \Gamma (b)}{\Gamma (a+b)}\,=\,\int_{0}^{1}d\alpha \alpha^{a-1}(1-\alpha)^{b-1}\,,
}
is the beta function. In this notation, the tensor structure of a three-point function can be written as
\es{threepointsum}{
	\mathcal{V}(P_{1},\polvec_{1};P_{2},\polvec_{2};P_{3},\polvec_{3})&=\sum_{I} C_{123}^{I}{\mathcal{V}}_I=\sum_{I}\tilde C_{123}^{I}\tilde{\mathcal{V}}_I\,,
}
for some constants $\tilde C^{I}$.
The reason for selecting this particular basis will become apparent shortly.

\subsection{Spinning three-point functions from geodesic Witten diagrams}

We now proceed to discuss how the same CFT spinning three-point functions can be obtained via particular bulk vertices and geodesic Witten diagrams.
We will begin with the simplest examples, where we will lay out some of the techniques and notation, and then move on to discuss the general structure of a three-point function computed via a bulk geodesic calculation.
Finally, we will derive the particular interaction vertices that produce the desired tensor structures.

As in the previous subsection, consider the simplest example: two scalar operators and one vector. 
The bulk calculation of this three-point function is generated by a bulk vertex between the dual AdS scalar fields, $\Phi_1(X),\Phi_2(X)$, and the dual vector field $\Phi_{3}^{(1)}(X,W)$. 
The vertex that connects these fields is essentially unique and is given by\footnote{We will denote bulk vertices by $V_{I}$, hopefully this will not be confused with the specific boundary tensor structures, $V_{i}$, or the general boundary structure $\mathcal{V}_{I}$.}
\begin{equation}
V_1 = \Phi_1(X) \nabla_A \Phi_2(X) \Phi_3^{A}(X) \, .
\end{equation}
We connect this vertex to boundary scalar operators at $P_1$ and $P_2$ using the scalar bulk-to-boundary propagators,
\es{bulkboundexpnotation}{
K_{\Delta_i}(P,X) &\equiv  \bbnorm{\Delta_i}{0}^{1/2} \langle O_{i}(P)\Phi(X)\rangle_{0} = \frac{\bbnorm{\Delta_i}{0}}{(-2 P \cdot X)^{\Delta_i}} \, .
}
Here, we have introduced some notation which will be useful in decluttering expressions later. We write the bulk-boundary correlator as a free two-point function or as an effective Wick contraction. We can do the same for the spin-1 bulk-boundary propagator \cite{Costa:2014kfa},
\es{spin1exp}{
K_{\Delta_i, \, 1}(P,\polvec;X,W) &\equiv \bbnorm{\Delta_i}{1}^{1/2} \langle O^{(1)}_{i}(P,\polvec)\Phi^{(1)}(X,W)\rangle_{0}  = \bbnorm{\Delta_i}{1} \frac{X \cdot C \cdot W }{(-2 P \cdot X)^{ \Delta_i+1}} \, 
}
where 
\begin{equation}
\bbnorm{\Delta}{\spin} = \frac{(\ell + \Delta -1)\Gamma(\Delta)}{2 \pi^{d/2} \Gamma(\Delta+1-h)} \, .
\end{equation}

\begin{figure}
	\centering
		\includegraphics[width=0.5\linewidth]{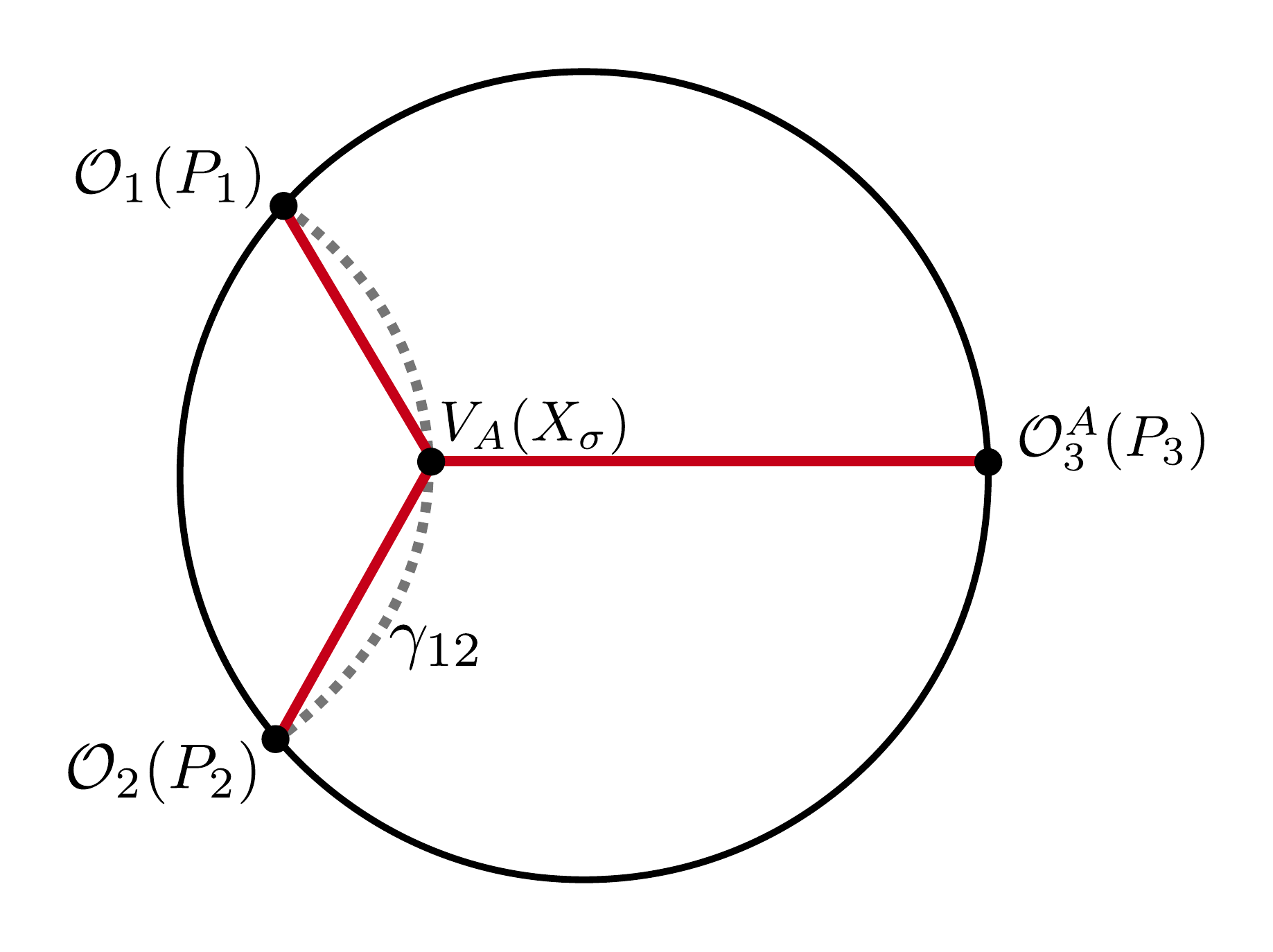}
	\caption{The geodesic witten diagram for two scalar and one vector operator. The vertex $V_{A}(X_{\sigma})$ is integrated over the geodesic $\gamma_{12}$.}
	\label{fig:gwd-ssv}
\end{figure}

We then consider not the standard Witten diagram computation of \cite{Freedman:1998tz}, but the geodesic computation where we restrict $X$ to lie on the geodesic connecting the boundary points $P_1$ and $P_2$, i.e. $X\to X_\s$. 
This geodesic three-point function is given by
\es{3ptexp}{
\mathcal{G}_1&=  \int_{\gamma_{12}} d\sigma \langle O_{1}(P_{1})O_{2}(P_{2})O_{3}^{(1)}(P_{3},\polvec_{3})V_{1}(X_\s)\rangle_{0}\\
&= \bbnorm{\Delta_1}{0}^{1/2}\bbnorm{\Delta_2}{0}^{1/2}\bbnorm{\Delta_3}{1}^{1/2} \int_{\gamma_{12}} d\sigma \left[K_{\Delta_1}(X_\s;P_{1}) \; \nabla K_{\Delta_2}(X_\s;P_{2})\cdot K_{\Delta_3, \, 1}(X_\s;\polvec_{3},P_{3})\right] \, .
}
Here we define the expectation value notation for multi-point correlators to be equal to the product of two-point functions, \`{a} la free Wick contractions. We depict this geodesic Witten computation in Figure \ref{fig:gwd-ssv}. 
To determine the integrand explicitly, note that
\begin{equation}
\nabla_A K_{\Delta}(P,X) = 2 \Delta \, K_{\Delta}(P,X) \, \frac{P_{A}}{-2 P \cdot X} \
\end{equation}
so that the integrand can be written
\begin{equation}
\mathcal{I}_1 = 2 b_1\left(\frac{1}{(-2 P_1 \cdot X_\s)^{\tau_1}(-2 P_2 \cdot X_\s)^{\tau_2}(-2 P_3 \cdot X_\s)^{\tau_3}} \right) \times \left( \frac{X_\s \cdot C_3 \cdot P_2}{-2 P_2 \cdot X_\s}\right) \; ,
\end{equation}
where the constant $b_1 = \Delta_2\bbnorm{\Delta_1}{0}^{1/2}\bbnorm{\Delta_2}{0}^{1/2}\bbnorm{\Delta_3}{1}^{1/2} $. The term in the second parenthesis looks similar to the tensor structure we found for  the scalar-scalar-vector correlator in Eq. \eqref{Vssv}. 
Indeed, we will show that it gives precisely this structure when integrated.

To compute the integral, note that two convenient parameterizations of the geodesic are given by
\bea \lab{geodsig}
X^A_\s &=&\frac{1}{\sqrt{(-2P_1\cdot P_2)}}[ e^{-\s}P_1^A + e^\s P_2^A]\\
X^A_\a &=& \frac{1}{\sqrt{(-2P_1\cdot P_2)}}\bigg[\sqrt{\frac{\a\,P_2\cdot P_3}{(1-\a)P_1\cdot P_3}} \,P_1^A +\sqrt{\frac{(1-\a)\,P_1\cdot P_3}{\a\,P_2\cdot P_3} }\,P_2^A\bigg]\,.\lab{geodalph}
\eea
where $\sigma$ is the proper length.\footnote{Equality of these two expressions gives the relation $e^\s =\sqrt{\frac{(1-\a)\,P_1\cdot P}{\a\,P_2\cdot P} }$,  and one easily computes the measure, obtaining for any function on the curve. 
\be
\int_{-\infty}^{+\infty} d\s f(X_\s) = \int_0^1 \frac{d\a}{2\a(1-\a)} f(X_\a)\,.
\ee}

This is just the intersection of the plane spanned by $\lbrace P_1,P_2 \rbrace$ with the AdS hyperbola $X^{2}=-1$ (written in the second case so that it remains invariant under rescalings of the $P_i$). 
It is then only a matter of plugging this in to Eq. \eqref{3ptexp} to find
\begin{equation}\label{eq:ssv-integral-exp}
\begin{split}
\mathcal{G}_1 &= b_1 \frac{V_{3,12}}{\left(P_{12}\right)^{\frac{\tau_1+\tau_2-\tau_3}{2}}\left(P_{13}\right)^{\frac{\tau _1+\tau _3-\tau _2}{2}}\left(P_{23}\right)^{\frac{\tau _2+\tau _3-\tau _1}{2}}}  \int_{0}^{1}d\alpha \alpha ^{\frac{\tau_{1}+\tau_{3}-\tau_{2}}{2} -1} (1-\alpha )^{\frac{\tau_{2}+\tau_{3}-\tau_{1}}{2}-1} \nonumber\\
 &= \tilde b_1 \langle O_{{1}}(P_{1})O_{{2}}(P_{2})O^{(1)}_{{3}}(P_{3})\rangle \, ,
\end{split}
\end{equation}
where $\tilde b_1 = b_1 B\left( \tfrac{\tau_{1}+\tau_{3}-\tau_{2}}{2},\tfrac{\tau_{2}+\tau_{3}-\tau_{1}}{2} \right) \times C_{123}^{-1}  $ and $B(x,y)$ is the standard Beta function. 
We have thus reproduced the boundary three-point function in Eq. \eqref{eq:spin-example-ssv} (up to an overall factor that depends on the operator normalization).

We can repeat this same procedure for a spin-$\spin$ bulk field. 
In this case the propagator is \cite{Costa:2014kfa}
\begin{equation}
K_{\Delta, \, \spin}(P_3,\polvec;X,W) = \bbnorm{\Delta}{\ell}\frac{(X \cdot C_3 \cdot W)^\spin }{(-2 P_3 \cdot X)^{ \Delta+\spin}} \, ,
\end{equation}
and the vertex is
\begin{equation}
V_\ell = \Phi_1(X) \nabla_{A_1} \ldots \nabla_{A_\spin}  \Phi_2(X) \Phi_3^{A_1 \ldots A_\spin}(X) \, .
\end{equation}
In an exactly analogous fashion to the above calculation, we find
\begin{equation}
\begin{split}
	\mathcal{G}_J &\propto \frac{(V_{3,12})^\ell}{\left(P_{12}\right)^{\frac{\tau_1+\tau_2-\tau_3}{2}}\left(P_{13}\right)^{\frac{\tau _1+\tau _3-\tau _2}{2}}\left(P_{23}\right)^{\frac{\tau _2+\tau _3-\tau _1}{2}}}  \int_{0}^{1}d\alpha \alpha ^{\frac{\tau_{1}+\tau_{3}-\tau_{2}}{2} -1} (1-\alpha )^{\frac{\tau_{2}+\tau_{3}-\tau_{1}}{2}-1} \nonumber\\
	&\propto  \langle O_{{1}}(P_{1})O_{{2}}(P_{2})O^{(\ell)}_{{3}}(P_{3})\rangle \, .
\end{split}
\end{equation}
This perfectly matches the result of the boundary calculation in Eq. \eqref{eq:spin-example-ssj}.

In the bulk, as in the dual CFT,  the situation becomes far nore complicated when there is no longer a unique bulk vertex or a unique boundary tensor structure. 
In the next section, we will introduce additional tools to organize the bulk calculation. The goal is to obtain a set of  bulk vertices for which geodesic integration produces a 1:1 relation with the boundary tensor structures of \reef{struc}.\footnote{Note, the vertices constructed here should not be confused with those written down in \cite{Sleight:2016dba} for the standard three point Witten diagrams integrated over the entire bulk.}

\subsubsection{A better bulk derivative}

Bulk calculations are complicated by the fact that, when derivatives act on the bulk-to-boundary propagators for particles with spin, they pick up two contributions:
\begin{equation}\label{eq:bad-derivative}
\langle O_i^{(\spin)}(P_i,\polvec ) \,  \tilde W \cdot \nabla \Phi^{(\ell)}_i(X,W) \rangle = \bbnorm{\Delta_i}{\spin}^{-1/2}K_{\Delta_i,\ell}(P_i,\polvec;X,W)\left( 2 \tau \frac{ \tilde W \cdot P_{i}}{(-2 P_i \cdot X)}  +  \ell \frac{ \tilde W \cdot C_{i} \cdot W }{(X \cdot C_i \cdot W )}\right) \, ,
\end{equation}
where we have introduced a new polarization vector, $\tilde{W}$, satisfying $\tilde{W}\cdot X=\tilde{W}^{2}=0$, to keep track of covariant derivative indices. 

Calculations would be much simpler were the second term eliminated. 
Thus, we will now design a new linear-operator that gives only the first term.  
Explicitly, given a spinning bulk field $\Phi^{(\ell)}_{i}(X,W)$, we will introduce a new degree-$m$ differential operator, $\diffop{\ell}{\Delta}{m}{\tilde W}{W}$ such that
\begin{equation}\label{eq:general-linear-op}
	\left\langle O_i^{(\ell)}(P_i,\polvec )\, \diffop{\ell}{\Delta_i}{m}{{\tilde W}}{W} \Phi_i^{(\ell)}(X,W)\right\rangle_0= \bbnorm{\Delta_i}{\spin}^{-1/2}\left(\frac{(\tilde{W}\cdot P_i)}{(-2 P_i \cdot X)} \right)^{m} \,  K_{\Delta_i,\ell}(P_i,\polvec;X,W) \,.
\end{equation}
To see how to construct $\diffop{\ell}{\Delta}{m}{\tilde W}{W}$, consider the case $m=1$. 
In this case, we need  to cancel the second term of \eqref{eq:bad-derivative}. We can do this by noting that the second terms can be generated by a combination of $X$ and $W$ derivatives:
\es{eq:extra-terms}{
	\left(\tilde W \cdot \nabla - \frac{1}{\ell}(W\cdot \nabla)(\tilde W \cdot \partial_W) \right)K_{\Delta,\ell}(P_i,\polvec;X,W) = 
	  \left( (\ell+1-\tau)  \frac{ \tilde W \cdot C_{i} \cdot W }{X \cdot C_i \cdot W}\right) K_{\Delta,\ell}(P_i,\polvec;X,W) \, .
}
Thus we can construct a derivative operator
\begin{equation}\label{eq:deriv-op}
\diffop{\ell}{\Delta}{1}{\tilde W}{W} = \frac{1}{2\tau}\left(\frac{\tau-1}{\tau-1 + \ell}\tilde W \cdot \nabla + \frac{1}{\tau-1+\ell}(W\cdot \nabla)(\tilde W \cdot \partial_W) \right) \, ,
\end{equation}
which, inside a bulk-to-boundary correlation function, gives the desired result:
\begin{equation}\label{sigmacovJdefexp}
\begin{split}
	\langle  O^{(\ell)}_i(P_i,\polvec)\, \diffop{\ell}{\Delta_i}{1}{\tilde W}{W} \Phi^{(\ell)}_{i}(\tilde{W},W;X)\rangle_0 &=\bbnorm{\Delta_i}{\spin}^{-1/2} \; \tilde{ W}\cdot P\frac{\left(X\cdot C\cdot W\right)^{\ell}}{(-2P\cdot X)^{\tau+1}}\,.
\end{split}
\end{equation}
The arbitrary $m$ operator is constructed as
\begin{equation}
\diffop{\ell}{\Delta}{m}{\tilde W}{W} = \prod_{k=0}^{m-1} \diffop{\ell}{\Delta+k}{1}{\tilde W}{W} \; .
\end{equation}
Although this operator may appear strange when written as an embedding space polynomial, it is quite standard. 
For example, the first derivative, when written out in index form, acts on an embedding polynomial as
\es{eq:for-ethan}{
	\mathbb{D}_{A}f_{\Delta,\spin}(X,W)&=\frac{\Pi_{A}^{A^{\prime}}}{2\tau}W^{B_{1}}\ldots W^{B_{\ell}}\left(\nabla_{A^{\prime}}f_{\Delta,\spin}(X)_{B_{1}\ldots B_{\spin}}-\frac{\spin}{\spin+\tau-1}\nabla_{[A^{\prime}}f_{\Delta,\spin}(X)_{B_{1}\ldots|B_{i}|\ldots B_{\spin}]}\right)\,.
}

\subsubsection{The general spinning bulk geodesic Witten diagram}\label{subsec:gen3pt}

Armed with our new operator, we are now in a position to write down the general vertex. 
Consider a vertex of the form 
\begin{equation}\label{eq:bulk-coupling}
\begin{split}
	{V}_{I}&=\Phi_{1}^{A_{1}\ldots A_{m_{1}}B_{1}\ldots B_{n_{12}}C_{1}\ldots C_{n_{13}}}\mathbb{D}^{(m_3)}\Phi^{D_{1}\ldots D_{m_{3}};E_{1}\ldots E_{m_{2}}\ \ \ \ \ \ \ \ \ F_{1}\ldots F_{n_{23}}}_{2 \ \ \ \ \ \ \ \ \ \ \ \ \ \ \ \ \ \ \ B_{1}\ldots B_{n_{12}}} \\
	&\;\;\;\;\;\quad \times \mathbb{D}^{(m_1+m_2)}\Phi_{3 A_{1}\ldots A_{m_{1}}E_{1}\ldots E_{m_{2}};D_{1}\ldots D_{m_{3}}C_{1}\ldots C_{n_{13}}F_{1}\ldots F_{n_{23}}}\; .
\end{split}
\end{equation}
We will then insert this vertex into a geodesic Witten correlator of the form
\begin{equation}\label{eq:bulk-geod-general}
\mathcal{G}_I = \int_{\gamma_{12}} d\sigma  \langle  O^{(\spin_1)}_1(P_1,\polvec)O^{(\spin_2)}_1(P_2,\polvec)O^{(\spin_3)}_1(P_3,\polvec)\, {V}_{I}(X_\sigma)  \rangle \, .
\end{equation}
The integrand of this equation is easy to compute by substituting  the differentiated propagators of  \eqref{eq:general-linear-op} and contracting indices as specified by the vertex in \eqref{eq:bulk-coupling}. 
One finds
\begin{equation}\label{eq:bulk-gen-integrand}
\begin{split}
	\mathcal{I}_{I}&= \langle  O^{(\spin_1)}_1(P_1,\polvec)O^{(\spin_2)}_1(P_2,\polvec)O^{(\spin_3)}_1(P_3,\polvec)\, {V}_{I}(X_\sigma)  \rangle\\
	&=b_I \left(\frac{X_{\sigma}\cdot C_{1}\cdot P_{3}}{-2P_{3}\cdot X_{\sigma}}\right)^{m_{1}}\left(\frac{X_{\sigma}\cdot C_{2}\cdot P_{3}}{-2P_{3}\cdot X_{\sigma}}\right)^{m_{2}}\left(\frac{X_{\sigma}\cdot C_{3}\cdot P_{2}}{-2P_{2}\cdot X_{\sigma}}\right)^{m_{3}}\\
	&\ \ \ \times\frac{\left(X_{\sigma}\cdot C_{12}\cdot X_{\sigma}\right)^{n_{12}}\left(X_{\sigma}\cdot C_{13}\cdot X_{\sigma}\right)^{n_{13}}\left(X_{\sigma}\cdot C_{23}\cdot X_{\sigma}\right)^{n_{23}}}{(-2 P_{1}\cdot X_{\sigma})^{\tau_{1}}(-2 P_{2}\cdot X_{\sigma})^{\tau_{2}}(-2 P_{3}\cdot X_{\sigma})^{\tau_{3}}}\,,
\end{split}
\end{equation}
where the constant $b_I =  \bbnorm{\Delta_1}{\spin_1}^{1/2}\bbnorm{\Delta_2}{\spin_2}^{1/2}\bbnorm{\Delta_3}{\spin_3}^{1/2} $.  

Our next task is to integrate this expression over the geodesic. 
Again, it is simply a matter of substituting in our geodesic parametrization to find
\begin{equation}\label{eq:three-points-gen}
\begin{split}
\mathcal{G}_I &=\frac{b_I/2}{4^{n_{12}+n_{13}+n_{23}}}\frac{V_{1,23}^{m_{1}}V_{2,13}^{m_{2}}V_{3,12}^{m_{3}}H_{12}^{n_{12}}H_{13}^{n_{13}}H_{23}^{n_{23}}}{d_{\tau_{1},\tau_{2},\tau_{3}}}\\
&\ \ \ \times \int d\alpha \alpha^{\kappa-1} (1-\alpha)^{\kappa^{\prime}-1}\left(1-8(1-\alpha)\frac{V_{1,23}V_{3,12}}{H_{13}}\right)^{n_{13}}\left(1+8\alpha\frac{V_{2,13}V_{3,12}}{H_{23}}\right)^{n_{23}}\\
\\
&=b_i/2\sum_{a=0}^{n_{13}}\sum_{b=0}^{n_{23}}\frac{f(\kappa,\kappa^{\prime};a,b)}{d_{\tau_{1},\tau_{2},\tau_{3}}}V_{1,23}^{m_{1}+a}V_{2,13}^{m_{2}+b}V_{3,12}^{m_{3}+a+b}H_{12}^{n_{12}}H_{13}^{n_{13}-a}H_{23}^{n_{23}-b} \; .
\end{split}
\end{equation}
The reader can now observe that the bulk vertex ${V}_I$ has given us exactly the boundary tensor structure $\tilde{\mathcal{V}}_I$ defined in \reef{struc}. 

We stress that the bulk vertices of \eqref{eq:bulk-coupling} are kinematic and need \emph{not} be thought of as coming from a consistent perturbative theory of higher spin fields in AdS spacetime. Instead, they are rather like effective field theory couplings which respect the symmetries of interest, in this case the linear action of the conformal group $SO(d+1,1)$ in embedding space.  Further these vertices are shown to generate  the complete set of boundary three-point structures only for geodesic Witten diagrams with the geodesic $\g_{12}$ between boundary points 1 and 2.

\section{Spinning Blocks from Bulk Diagrams}\label{sec:spinning-blocks}

\begin{figure}
	\centering
		\includegraphics[width=0.6\linewidth]{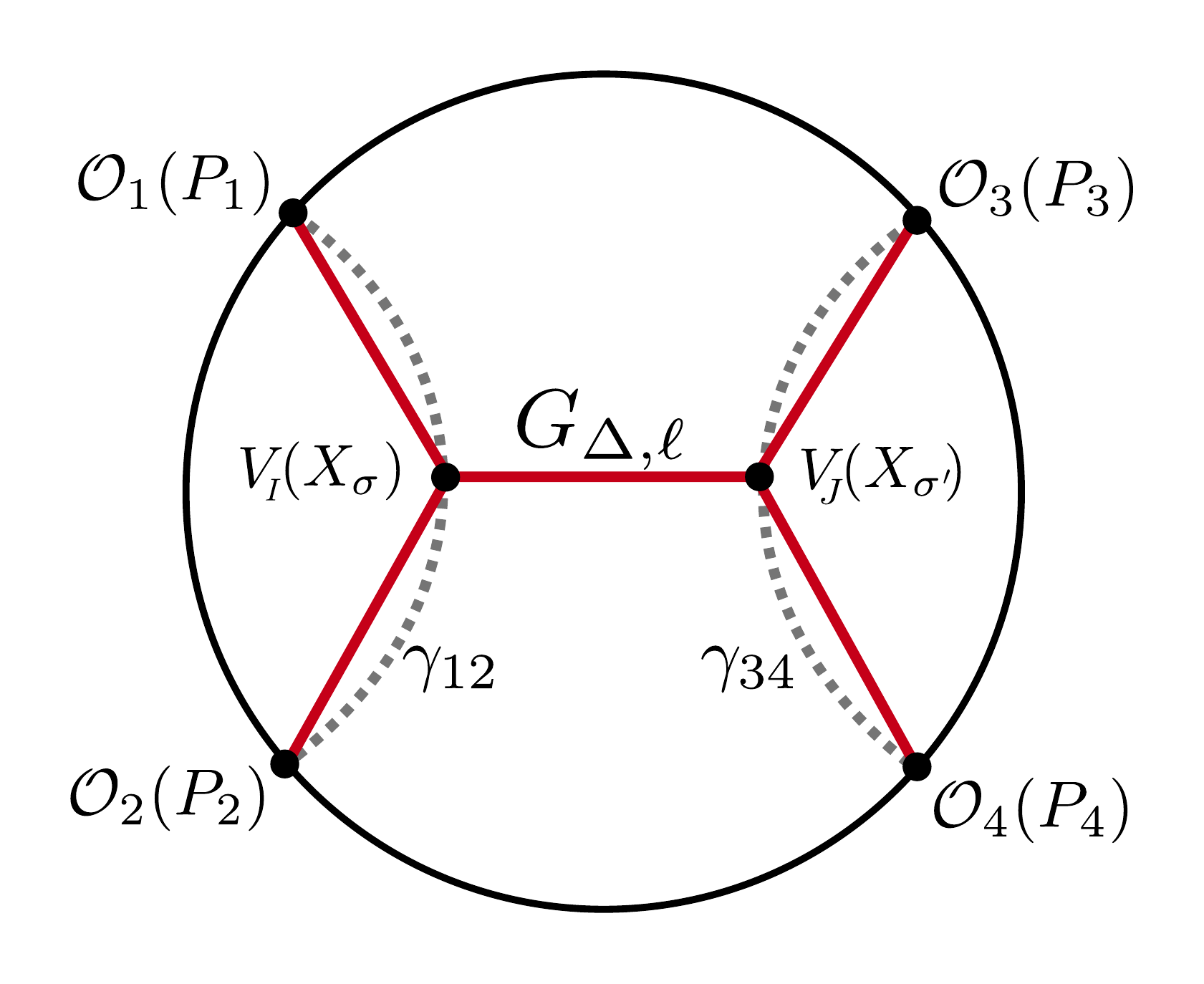}
	\caption{A geodesic witten diagram for a spinning conformal block. The vertices $V_{I}$ and $V_{J}$ are integrated over the geodesics $\gamma_{12}$ and $\gamma_{34}$ respectively.}
	\label{fig:gwd-spinning}
\end{figure}

Equipped with our dictionary relating three-point bulk couplings and boundary structures, we are now in a position to write down an expression for the spinning conformal block as a GWD:

\es{geodesicblock}{
\hat{\bbprop}_{\{\Delta_{i},\spin_{i}\};\Delta,\spin}^{IJ}(P_i)&=\int_{\gamma_{12}} d\sigma \int_{\gamma_{13}}d\sigma^{\prime}\langle O_{1}(P_{1},\polvec_{1})O_{2}(P_{2},\polvec_{2})V^{I}(X_{\sigma})V^{J}(X_{\sigma^{\prime}})O_{3}(P_{3},\polvec_{3})O_{4}(P_{4},\polvec_{4})\rangle_{0}\,.
}
The GWD that is described by the correlator on the right hand side is pictured in Figure \ref{fig:gwd-spinning}.
The  subscript $\>_0$ indicates Wick contractions, as in  \reef{bulkboundexpnotation},  that give bulk-boundary propagators, as well as the contraction $\<\Phi_3(X_\s)\Phi_3(X_{\s'}\>_0$  that produces a bulk-bulk propagator.

This presentation of the spinning conformal block is our main result.   
To verify this claim, in the remainder of this section we will establish that our GWD expression satisfies the conformal Casimir equation and has the correct short distance behavior. Finally, we will use the split representation of the bulk-to-bulk propagator, to relate this GWD formula for the block to the shadow integral formalism of \cite{SimmonsDuffin:2012uy}. 

Before moving on, however, we pause to unpack expression \eqref{geodesicblock} which is a compact representation of a relatively elaborate object in general. As a concrete example, let's consider the case
$\ell_1=1;~\ell_2=\ell_3=\ell_4=0;~\ell=1$ where there is both an external and exchanged operator with spin:

\paragraph{Illustration of (4.1) for blocks  with $\ell_1=1;~\ell_2=\ell_3=\ell_4=0;~\ell=1$:}

In this case we have two possible structures for the first vertex, $(\{n_{13}=1,n_{12}=n_{23}=m_{i}=0\},\{m_{1}=m_{3}=1,m_{2}=n_{ij}=0\})$ and a unique structure for the second. (The integers $n_{ij},~m_i$ are
defined in \reef{oldbasis}.)
The first two structures are encoded through the bulk vertices at $X$,
\es{bulkvertex1}{
V_{1}(X)&=\Phi_{1}(X)^{A}\Phi_{2}(X)\Phi(X)_{A}\\
V_{2}(X)&=\Phi_{1}(X)^{A}\mathbb{D}_{B}\Phi_{2}(X)\mathbb{D}_{A}\Phi(X)^{B}\\
&=\frac{1}{2\tau(\tau-2)}\Phi_{1}(X)^{A}\nabla_{B}\Phi_{2}(X)\left((\tau-1)\nabla_{A}\Phi(X)^{B} - \nabla_{B}\Phi(X)^{A}\right)
}
The unique second structure is given by a vertex at $X^{\prime}$
\es{bulkvertex2}{
V(X^{\prime})&=\Phi_{3}(X^{\prime})\mathbb{D}_{A^{\prime}}\Phi_{4}(X^{\prime})\Phi(X^{\prime})^{A^{\prime}}\\
&=\Phi_{3}(X^{\prime})\nabla_{A^{\prime}}\Phi_{4}(X^{\prime})\Phi(X^{\prime})^{A^{\prime}}\,.
}
Putting these together gives the two possible blocks. The first is given by,
\es{geodesicblockex1}{
\hat{\bbprop}_{\{\Delta_{i},\spin_{i}\};\Delta,1}^{1}(P_{i})&=\int_{\gamma_{12}} d\sigma \int_{\gamma_{13}}d\sigma^{\prime}\langle O_{1}(P_{1},\polvec_{1})O_{2}(P_{2},\polvec_{2})V_{1}(X_{\sigma})V(X_{\sigma^{\prime}})O_{3}(P_{3},\polvec_{3})O_{4}(P_{4},\polvec_{4})\rangle_{0}\\
&=\int_{\gamma_{12}} d\sigma \int_{\gamma_{13}}d\sigma^{\prime}K(P_{1},X_{\sigma})^{A}K(P_{2},X_{\sigma})\bbprop(X_{\sigma},X_{\sigma^{\prime}})_{AA^{\prime}}K(P_{3},X_{\sigma^{\prime}})\mathbb{D}^{\prime A^{\prime}}K(P_{4},X_{\sigma^{\prime}})\\
&=\int_{\gamma_{12}} d\sigma \int_{\gamma_{13}}d\sigma^{\prime}K(P_{1},X_{\sigma})^{A}K(P_{2},X_{\sigma})\bbprop(X_{\sigma},X_{\sigma^{\prime}})_{AA^{\prime}}K(P_{3},X_{\sigma^{\prime}})\nabla^{\prime A^{\prime}}K(P_{4},X_{\sigma^{\prime}})\,.
}
The second by,
\es{geodesicblockex2}{
\hat{\bbprop}_{\{\Delta_{i},\spin_{i}\};\Delta,1}^{2}(P_{i})&=\int_{\gamma_{12}} d\sigma \int_{\gamma_{13}}d\sigma^{\prime}\langle O_{1}(P_{1},\polvec_{1})O_{2}(P_{2},\polvec_{2})V_{2}(X_{\sigma})V(X_{\sigma^{\prime}})O_{3}(P_{3},\polvec_{3})O_{4}(P_{4},\polvec_{4})\rangle_{0}\\
&=\int_{\gamma_{12}} d\sigma \int_{\gamma_{13}}d\sigma^{\prime}K(P_{1},X_{\sigma})^{A}\mathbb{D}_{B}K(P_{2},X_{\sigma})\mathbb{D}_{A}\bbprop(X_{\sigma},X_{\sigma^{\prime}})^{B}_{A^{\prime}}K(P_{3},X_{\sigma^{\prime}})\mathbb{D}^{\prime A^{\prime}}K(P_{4},X_{\sigma^{\prime}})\\
&=\frac{\tau-1}{2\tau(\tau-2)}\int_{\gamma_{12}} d\sigma \int_{\gamma_{13}}d\sigma^{\prime}K(P_{1},X_{\sigma})^{A}\nabla_{B}K(P_{2},X_{\sigma})\nabla_{A}\bbprop(X_{\sigma},X_{\sigma^{\prime}})^{B}_{A^{\prime}}K(P_{3},X_{\sigma^{\prime}})\nabla^{\prime A^{\prime}}K(P_{4},X_{\sigma^{\prime}})\\
&-\frac{1}{2\tau(\tau-2)}\int_{\gamma_{12}} d\sigma \int_{\gamma_{13}}d\sigma^{\prime}K(P_{1},X_{\sigma})^{A}\nabla_{B}K(P_{2},X_{\sigma})\nabla_{B}\bbprop(X_{\sigma},X_{\sigma^{\prime}})^{A}_{A^{\prime}}K(P_{3},X_{\sigma^{\prime}})\nabla^{\prime A^{\prime}}K(P_{4},X_{\sigma^{\prime}})\,.
}

\paragraph{General Case} To write out the integrand for a generic block, we need the full machinery of Section \ref{subsec:gen3pt}. 
In this case, the integrand, written out explicitly, is given by
\es{integrand}{
\mathcal{I}_{IJ}(\sigma,\sigma^{\prime})&=K(P_{1},X_{\sigma})^{A_{1}\ldots A_{m_{1}}B_{1}\ldots B_{n_{12}}C_{1}\ldots C_{n_{13}}}\mathbb{D}^{(m_3)}K(P_{2},X_{\sigma})^{D_{1}\ldots D_{m_{3}};E_{1}\ldots E_{m_{2}}\ \ \ \ \ \ \ \ \ F_{1}\ldots F_{n_{23}}}_{2 \ \ \ \ \ \ \ \ \ \ \ \ \ \ \ \ \ \ \ B_{1}\ldots B_{n_{12}}}\\
&\times\mathbb{D}^{\prime(m_1^{\prime}+m_2^{\prime})}\mathbb{D}^{(m_1+m_2)}\bbprop_{A_{1}\ldots A_{m_{1}}E_{1}\ldots E_{m_{2}}A^{\prime}_{1}\ldots A^{\prime}_{m^{\prime}_{1}}E^{\prime}_{1}\ldots E^{\prime}_{m^{\prime}_{2}};D_{1}\ldots D_{m_{3}}C_{1}\ldots C_{n_{13}}F_{1}\ldots F_{n_{23}}D^{\prime}_{1}\ldots D^{\prime}_{m^{\prime}_{3}}C^{\prime}_{1}\ldots C^{\prime}_{n^{\prime}_{13}}F_{1}\ldots F^{\prime}_{n^{\prime}_{23}}}\\
&\times K(P_{3},X_{\sigma^{\prime}})^{A^{\prime}_{1}\ldots A^{\prime}_{m_{1}^{\prime}}B^{\prime}_{1}\ldots B^{\prime}_{n_{12}^{\prime}}C^{\prime}_{1}\ldots C^{\prime}_{n_{13}^{\prime}}}\mathbb{D}^{\prime(m_3^{\prime})}K(P_{4},X_{\sigma^{\prime}})^{D^{\prime}_{1}\ldots D^{\prime}_{m_{3}^{\prime}};E^{\prime}_{1}\ldots E^{\prime}_{m_{2}^{\prime}}\ \ \ \ \ \ \ \ \ F^{\prime}_{1}\ldots F^{\prime}_{n_{23}^{\prime}}}_{2 \ \ \ \ \ \ \ \ \ \ \ \ \ \ \ \ \ \ \ B^{\prime}_{1}\ldots B^{\prime}_{n_{12}^{\prime}}}\,.
}
and the block itself comes from integrating over the pair of geodesics:
\es{expandblock}{
\hat{\bbprop}_{\{\Delta_{i},J_{i}\};\Delta,J}^{IJ}(P_{i})&=\int_{\gamma_{12}} d\sigma \int_{\gamma_{13}}d\sigma^{\prime}\mathcal{I}_{IJ}(\sigma,\sigma^{\prime})\,.
}
\subsection{Casimir Equation}
Here, we demonstrate that the spinning GWD presentation of the block, \eqref{geodesicblock}, satisfies the appropriate Casimir equation. The argument is similar  to the argument for scalar external operators presented in \cite{Hijano:2015zsa}. The essential point is that our construction of the spinning block transforms as a scalar under embedding space rotations. This combined  with the fact that the bulk-to-bulk propagator is an eigenfunction of the bulk Laplacian ensures that the block satisfies the correct Casimir equation. In this subsection we spell out the details of this argument.

Conformal invariance of a correlation function can be presented as the vanishing of commutators involving the conformal generators: 
\es{confinv}{
\sum_{i=1}^{N}\langle O_{1}(P_{1};\polvec_{1})O_{2}(P_{2};\polvec_{2})\ldots\left[L_{AB},O_{i}(P_{i};\polvec_{i})\right]\ldots O_{N}(P_{N};\polvec_{N})\rangle&=0\,,
}
or equivalently,
\es{confinvdiff}{
\sum_{i=1}^{N}\mathcal{L}_{AB}^{i}\langle O_{1}(P_{1};\polvec_{1})O_{2}(P_{2};\polvec_{2})\ldots O_{N}(P_{N};\polvec_{N})\rangle&=0\,.
}
Here, $\mathcal{L}_{AB}^{i}=P_{i A}\frac{\partial}{\partial P_{i}^{B}}-P_{i B}\frac{\partial}{\partial P_{i}^{A}}+\polvec_{i A}\frac{\partial}{\partial \polvec_{i}^{B}}-\polvec_{i B}\frac{\partial}{\partial \polvec_{i}^{A}}$, is the differential operator implementing conformal transformations. The main advantage of the embedding space formalism we have been using is that these conformal transformations are represented as $SO(d+1,1)$ rotations.

With this notation under our belts, we can write down the Casimir equation satisfied by the conformal block,
\es{spinningCas}{
\frac{1}{2}(\mathcal{L}_{AB}^{1}+\mathcal{L}_{AB}^{2})^{2}\hat{\bbprop}_{\{\Delta_{i},\spin_{i}\};\Delta,j}^{IJ}(P_i)&=-C_{2}(\Delta,\spin)\hat{\bbprop}_{\{\Delta_{i},\spin_{i}\};\Delta,j}^{IJ}(P_i)\,.
}
The Casimir eigenvalue is $C_{2}(\Delta,\spin)=\Delta(\Delta-d)+\spin(\spin+d-2)$.

\begin{figure}
	\centering
	\includegraphics[width=0.5\linewidth]{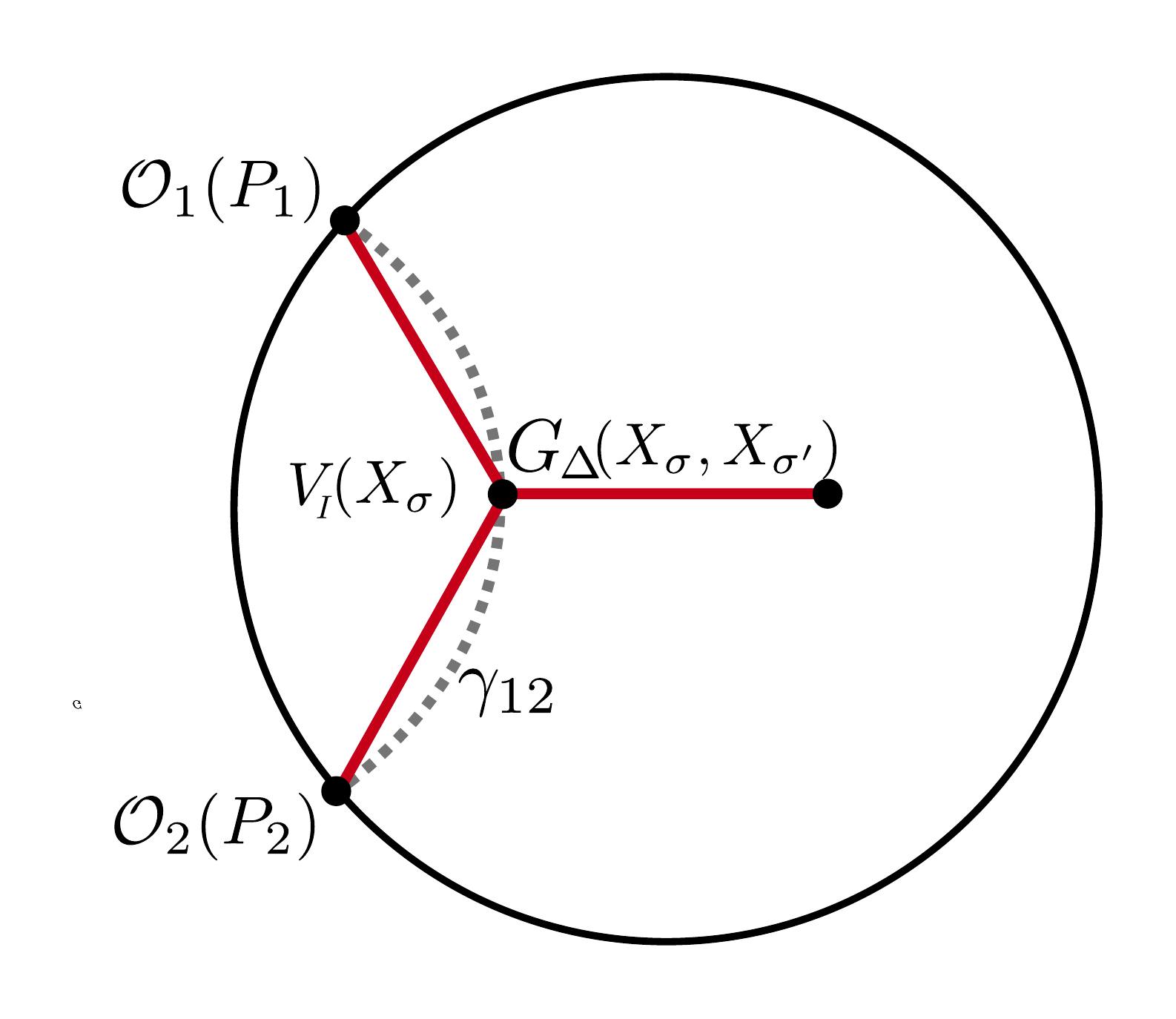}
	\caption{The function $F^I$ represents `half' of the geodesic Witten diagram.}
	\label{fig:f-object}
\end{figure}

We would like to show that the GWD presentation of the spinning conformal block satisfies the Casimir equation, \eqref{spinningCas}. To see this, let's focus on the piece of the GWD that contains dependance on $P_{1}$ and $P_{2}$. If we introduce,
\es{Fdef}{
F^{I}(P_{1},\polvec_{1};P_{2},\polvec_{2};X_{\sigma^{\prime}},W^{\prime})&
\equiv\int_{\gamma_{12}} d\sigma \langle O_{1}(P_{1},\polvec_{1}) O_{2}(P_{2},\polvec_{2})V^{I}(X_{\sigma})\Phi(X_{\sigma^{\prime}},W^{\prime})\rangle_{0}\,,\\
}
(which we depict in Figure \ref{fig:f-object}) then the full block can be written as,
\es{blockfromF}{
\hat{\bbprop}_{\{\Delta_{i},J_{i}\};\Delta,j}^{IJ}(u,v)=
\int_{\gamma_{34}} d\sigma^{\prime} &K(P_{3},\polvec_{3};X_{\sigma^{\prime}})
\mathbb{D}^{\prime}K(P_{4},\polvec_{4};X_{\sigma^{\prime}})
\mathbb{D}^{\prime}F^{I}(P_{1},\polvec_{1};P_{2},\polvec_{2};X_{\sigma^{\prime}})\,,
}
where the tensor and derivative indices have been suppressed. They are contracted as in equations \eqref{integrand} and \eqref{expandblock}. 

The object, $F^{I}$, is invariant under simultaneous rotations of all the arguments,
\es{inveqF}{
\left(\mathcal{L}_{AB}^{1}+\mathcal{L}_{AB}^{2}+\mathcal{L}_{AB}^{\prime}\right)F^{I}(P_{1},\polvec_{1};P_{2},\polvec_{2};X_{\sigma^{\prime}},W^{\prime})&=0\,,
}
and thus it is easy to compute the action of the Casimir on $F$.
\es{CasF}{
\frac{1}{2}\left(\mathcal{L}_{AB}^{1}+\mathcal{L}_{AB}^{2}\right)^{2}F^{I}(P_{1},\polvec_{1};P_{2},\polvec_{2};X_{\sigma^{\prime}},W^{\prime})&=\frac{1}{2}\left(\mathcal{L}_{AB}^{\prime}\right)^{2}F^{I}(P_{1},\polvec_{1};P_{2},\polvec_{2};X_{\sigma^{\prime}},W^{\prime})\\
&=-\left[\nabla^{\prime2}+\spin (\spin+d-1)\right]F^{I}(P_{1},\polvec_{1};P_{2},\polvec_{2};X_{\sigma^{\prime}},W^{\prime})\\
&=-C_{2}(\Delta,\spin)F^{I}(P_{1},\polvec_{1};P_{2},\polvec_{2};X_{\sigma^{\prime}},W^{\prime})\,.
}
In the second line we used the fact (see Appendix~\ref{app:cas} or \cite{Pilch:1984xx}) that the bulk Casimir is simply related to the bulk Laplacian, $\left(\nabla^{2}+\spin(d+\spin-1)\right)=C_{2}(\Delta,\spin)$ when acting on harmonic tensors. In the third line, we used the action of the Laplacian on the bulk-to-bulk propagator of the exchanged field, which is implicit in \eqref{Fdef}.

This is progress, but it does not yet establish the Casimir equation for the block. In the block, $F^{I}$ appears with additional derivatives acting on it, and so to complete the proof that the expression \eqref{geodesicblock} satisfies the Casimir equation, we must establish that,
\es{commrel}{
(\mathcal{L}_{AB}^{\prime})^{2}\mathbb{D}^{\prime m}F^{I}(P_{1},\polvec_{1};P_{2},\polvec_{2};X^{\prime},W^{\prime})_{A_{1},\ldots,A_{m}}&=\mathbb{D}^{\prime m}(\mathcal{L}_{AB}^{\prime})^{2}F^{I}(P_{1},\polvec_{1};P_{2},\polvec_{2};X^{\prime},W^{\prime})_{A_{1}\ldots A_{m}}\,.
}
This can be shown by explicit calculation and is analogous to the the boundary statement, that the Casimir in the boundary CFT commutes with boundary derivatives.\footnote{To see this it is convenient to adopt index free notation by introducing new polarization vectors, $\{\tilde{W}_{i}\}$ for each derivative in $\mathcal{D}$. By recursively applying the identity,
\es{Lactid}{
\mathcal{L}^{\prime}_{AB}\left[\tilde{W}^{C}_{N}\partial^{\prime}_{C}f(\{\tilde{W}_{i};X_{\sigma},W;X_{\sigma^{\prime}},W^{\prime})\right]&=\tilde{W}^{C}\partial^{\prime}_{C}\mathcal{L}^{\prime}_{AB}\left[f(\{\tilde{W}_{i}\};X_{\sigma},W;X_{\sigma^{\prime}},W^{\prime})\right]\,,
}
to move $\mathcal{L}^{\prime}_{AB}=\mathcal{L}^{(X_{\sigma^{\prime}})}_{AB}+\mathcal{L}^{(\tilde{W})}_{AB}+\mathcal{L}^{(W^{\prime})}_{AB}$ (the total rotation generator) through each derivative, we arrive at the desired result. The single derivative identity can be verified by direct computation, using the definitions of the various $\mathcal{L}_{AB}$. 
}  

There is also a simple heuristic argument, which shows that $\mathcal{L}_{AB}\mathbb{D}^{m}\bbprop(X,W;X^{\prime},W^{\prime})_{A_{1}\ldots A_{m}}=\mathbb{D}^{m}\mathcal{L}_{AB}\bbprop(X,W;X^{\prime},W^{\prime})_{A_{1}\ldots A_{m}}$, and thus the corresponding statement for $F^{I}$. 
The bulk-to-bulk propagator is a bulk two point function of free field operators, ie. $\bbprop(X,W;X^{\prime},W^{\prime})=\langle\Phi(X,W)\Phi(X^{\prime},W^{\prime})\rangle_{0}$.
Since $(\mathcal{L}_{AB}^{\prime})^{2}$ acts only on these operators, we can write
\es{heuristic}{
\mathcal{L}_{AB}\mathbb{D}^{m}\bbprop(X,W;X^{\prime},W^{\prime})_{A_{1}\ldots A_{m}}&=\langle[L_{AB},\mathbb{D}^{m}\Phi(X,W)_{A_{1}\ldots A_{m}}]\Phi(X^{\prime},W^{\prime})\rangle_{0}\\
&=\left(\mathbb{D}^{m}\langle[L_{AB},\Phi(X,W)]\Phi(X^{\prime},W^{\prime})\rangle_{0}\right)_{A_{1}\ldots A_{m}}\\
&=\mathbb{D}^{m}\mathcal{L}_{AB}\bbprop(X,W;X^{\prime},W^{\prime})_{A_{1}\ldots A_{m}}\,.
}
Trusting this heuristic, or relying on the explicit derivation, we now see that the spinning GWD satisfies the Casimir equation, 
\es{CasblockfromF}{
-\frac{1}{2}\left(\mathcal{L}_{AB}^{1}+\mathcal{L}_{AB}^{2}\right)^{2}&\hat{\bbprop}_{\{\Delta_{i},\spin_{i}\};\Delta,\spin}^{IJ}(u,v)
=C_{2}(\Delta,\spin)\hat{\bbprop}_{\{\Delta_{i},\spin_{i}\};\Delta,\spin}^{IJ}(u,v)\,.
}
This is the first step in establishing the GWD presentation of the block.
\subsection{Shadows and Short Distances}\label{sec:shadows-short}
To finish the argument that the expression, \eqref{geodesicblock}, gives the spinning conformal block, we need to show that it has the correct behavior in the $u=\frac{P_{12}P_{34}}{P_{13}P_{24}}\rightarrow0$ limit. We will actually go further, and explain how the spinning GWD presentation of the block is equivalent to the shadow integral representation of \cite{SimmonsDuffin:2012uy}. The connection is almost identical to that described for the scalar case in Section~\ref{sec: shadow1}, but now decorated with spin. 

\begin{figure}
	\centering
	\includegraphics[width=0.9\linewidth]{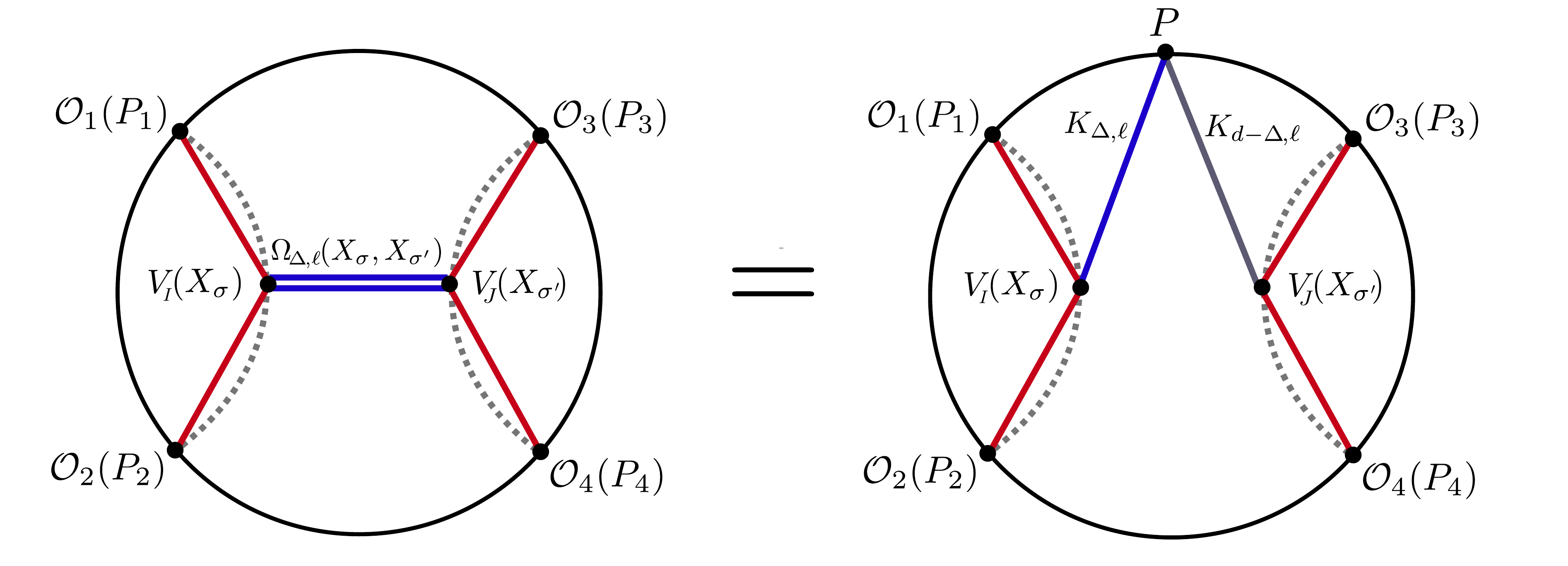}
	\caption{The geodesic Witten diagram for the spinning harmonic function. On the LHS, the bulk propagator in the regular spinning GWD has been replaced by a bulk harmonic, represented by the double line. On the RHS, an equivalent representation for the bulk harmonic is given in the `split representation' where bulk-to-boundary operators for the bulk field and its shadow are integrated over the boundary, as in Eq. \eqref{Omegwspin}.}
	\label{fig:harmonic-bulk-spinning}
\end{figure}

The spinning bulk harmonic function of spin $\spin$ and dimension $\Delta$ can be written as a difference of bulk-to-bulk propagators of dimension $\Delta$ and $d-\Delta$. 
\es{Omegwspin}{
\Omega_{\Delta,\spin}(X,W;X^{\prime},W^{\prime})&=\bbprop_{\Delta,\spin}(X,W;X^{\prime},W^{\prime})-\bbprop_{d-\Delta,\spin}(X,W;X^{\prime},W^{\prime})\,.
}
Integrating this expression over the two geodesics, $\gamma_{12}$ and $\gamma_{34}$ connecting $P_{1}$ to $P_{2}$ and $P_{3}$ to $P_{4}$ respectively (see Figure \ref{fig:harmonic-bulk-spinning}), thus gives a sum of direct and shadow GWDs,
\es{bothblocksfromom}{
&\int_{\gamma_{12}} d\sigma \int_{\gamma_{13}}d\sigma^{\prime}K_{\Delta_{1},\spin_{1}}(P_{1},U_{1};X_{\sigma})\mathbb{D}K_{\Delta_{2},\spin_{2}}(P_{2},U_{2};X_{\sigma})\mathbb{D}\mathbb{D}^{\prime}\Omega_{\Delta,\spin}(X_{\sigma};X_{\sigma^{\prime}})K_{\Delta_{3},\spin_{3}}(P_{3},U_{3};X_{\sigma^{\prime}})\mathbb{D}^{\prime}K_{\Delta_{4},\spin_{4}}(P_{4},U_{4};X_{\sigma^{\prime}})\\
&=\hat{\bbprop}_{\{\Delta_{i},\spin_{i}\};\Delta,\spin}^{IJ}-\hat{\bbprop}_{\{\Delta_{i},\spin_{i}\};d-\Delta,\spin}^{IJ}\,.
}
Index contraction are again suppressed as in \eqref{blockfromF}.

The two propagators in \eqref{bothblocksfromom} can be distinguished by their branch structure around $-X\cdot X^{\prime}\rightarrow\infty$,
\es{largedistlimit}{
\bbprop_{\Delta,\spin}(X,W;X^{\prime},W^{\prime})&\sim \frac{\left(2(W\cdot X^{\prime})(W^{\prime}\cdot X)-2(X\cdot X^{\prime})(W\cdot W^{\prime})\right)^{\spin}}{(-2X\cdot X^{\prime})^{\Delta+\spin}} \; .
}
We can thus use, a bulk projector, $\mathcal{P}_{\Delta}$, to write the bulk-to-bulk propagator for an operator without its shadow. 
\es{projectwspin}{
\bbprop_{\Delta,\spin}(X,W;X^{\prime},W^{\prime})&=\mathcal{P}_{\Delta}\Omega_{\Delta,\spin}(X,W;X^{\prime},W^{\prime})\,,
}
and similarly the GWD can be written as,
\es{blockint}{
\hspace{-.1\textwidth}& \hspace{.1\textwidth} \hat{\bbprop}_{\{\Delta_{i},\spin_{i}\};\Delta,\spin}^{IJ}(P_{i})=\\
\hspace{-.1\textwidth}&\int_{\gamma_{12}} d\sigma \int_{\gamma_{13}}d\sigma^{\prime}K_{\Delta_{1},\spin_{1}}(P_{1},U_{1};X_{\sigma})\mathbb{D}K_{\Delta_{2},\spin_{2}}(P_{2},U_{2};X_{\sigma})\mathcal{P}_{\Delta}\mathbb{D}\mathbb{D}^{\prime}\Omega_{\Delta,\spin}(X_{\sigma};X_{\sigma^{\prime}})K_{\Delta_{3},\spin_{3}}(P_{3},U_{3};X_{\sigma^{\prime}})\mathbb{D}^{\prime}K_{\Delta_{4},\spin_{4}}(P_{4},U_{4};X_{\sigma^{\prime}})\,.
}

To make contact with the shadow formalism, we can rewrite this using the split representation of the bulk harmonic function,
\es{expforomega}{
\Omega_{\Delta,\spin}(X,W;X^{\prime},W^{\prime})&=\int d^{d}PK_{\Delta,\spin}(P;X,W)^{\{A\}}K_{d-\Delta,\spin}(P;X^{\prime},W^{\prime})_{\{A\}}\\
&=\int d^{d}Pd^{d}P^{\prime}K_{\Delta,\spin}(P;X,W)^{\{A\}}T_{\{A\}\{B\}}(P,P^{\prime})K_{\Delta,\spin}(P^{\prime};X^{\prime},W^{\prime})^{\{B\}}\,,
}
where we have introduced the tensor structure, $T_{\{A\}\{B\}}$ needed to convert from the shadow propagator back to the direct, see \eqref{dirshadJ}, and used a multi-index notation $\{\cdot\}$.

Plugging this into \eqref{bothblocksfromom} and performing the bulk geodesic integrals, gives the expression,
\es{shadowint}{
\hat{\Omega}_{\Delta,\spin}(P_{i},\polvec_{i})&=\int d^{d}Pd^{d}P^{\prime}\langle O_{1}(P_{1},\polvec_{1})O_{2}(P_{2},\polvec_{2})O_{\Delta,\spin}^{\{A\}}(P)\rangle_{I} T_{\{A\}\{B\}}(P,P^{\prime})\langle O_{\Delta,\spin}^{\{B\}}(P) O_{3}(P_{3},\polvec_{3})O_{4}(P_{4},\polvec_{4})\rangle_{J} \, ,
}
which is now formulated in terms of CFT quantities only. 
As in the scalar case, this harmonic function is the sum of two conformal blocks, the direct and shadow, which are distinguished by their branch structure around the OPE limit, $u=0$. To single out the direct block, we must use a \emph{boundary} projector to select the correct behavior around $u\rightarrow0$.
\es{blockagain}{
\hat{\bbprop}_{\{\Delta_{i},\spin_{i}\};\Delta,\spin}^{IJ \; (CFT)}(P_{i},U_{i})&=\mathcal{P}_{\Delta}\hat{\Omega}_{\Delta,\spin}(P_{i},\polvec_{i})\,.
}
where we have used the `CFT' superscript to emphasize that this is the true CFT conformal block, rather than the yet to be equated GWD. 
This formulation of the block, equations \eqref{shadowint} and \eqref{blockagain}, as a boundary projection of the integral of two three-point functions is exactly the shadow integral representation of the block used in \cite{SimmonsDuffin:2012uy}.\footnote{Our equation \eqref{shadowint} corresponds to (4.4) and (4.6) in \cite{SimmonsDuffin:2012uy}, with our $I,J$ indices mapping to $m,n$ indices there.}

\begin{figure}
	\centering
	\includegraphics[width=0.7\linewidth]{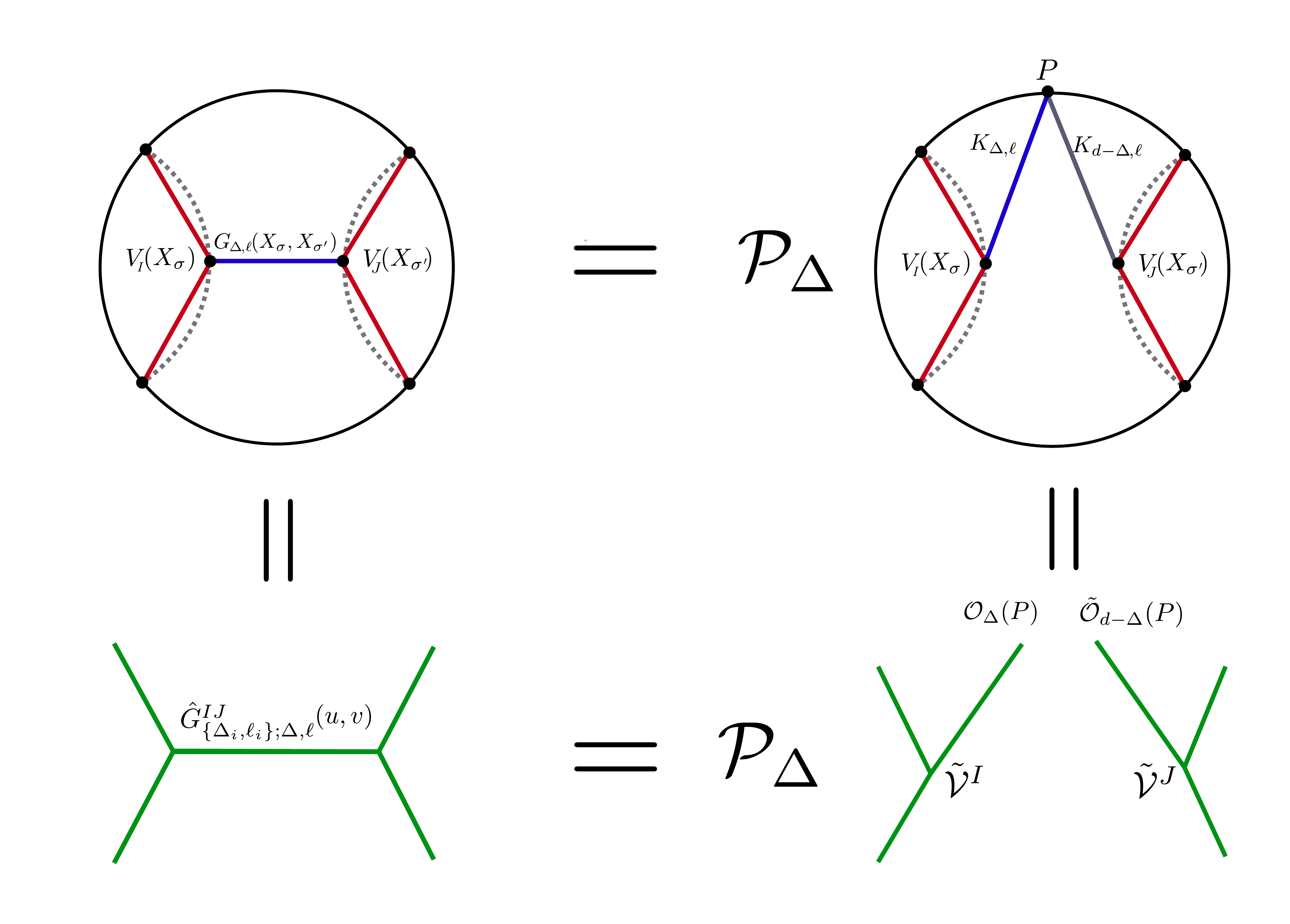}
	\caption{As in the scalar case, the spinning geodesic Witten diagram is recovered from acting on the bulk harmonic with a suitable projector. This is equivalent to acting with a projector on the shadow-operator representation of the harmonic function, and demonstrates the connection between the spinning conformal block and the spinning GWD.}
	\label{fig:bulk-proj-spinning}
\end{figure}

The equivalence between the GWD formulation and the shadow integral representation is, as in the scalar case, the equivalence of bulk and boundary projectors (as depicted diagrammatically in Figure \ref{fig:bulk-proj-spinning}). Projecting onto the correct bulk structure in the $-X\cdot X^{\prime}\rightarrow\infty$ limit maps precisely onto the boundary projector selecting the direct block:
\es{blockint2}{
\hspace{-.1\textwidth}&  \hat{\bbprop}_{\{\Delta_{i},\spin_{i}\};\Delta,\spin}^{IJ \; (CFT)}(P_{i},U_{i}) = \mathcal{P}_{\Delta}\hat{\Omega}_{\Delta,\spin}(P_{i},\polvec_{i})=\\
\hspace{-.1\textwidth}&\int_{\gamma_{12}} d\sigma \int_{\gamma_{13}}d\sigma^{\prime}K_{\Delta_{1},\spin_{1}}(P_{1},U_{1};X_{\sigma})\mathbb{D}K_{\Delta_{2},\spin_{2}}(P_{2},U_{2};X_{\sigma})\mathcal{P}_{\Delta}\mathbb{D}\mathbb{D}^{\prime}\Omega_{\Delta,\spin}(X_{\sigma};X_{\sigma^{\prime}})K_{\Delta_{3},\spin_{3}}(P_{3},U_{3};X_{\sigma^{\prime}})\mathbb{D}^{\prime}K_{\Delta_{4},\spin_{4}}(P_{4},U_{4};X_{\sigma^{\prime}})\\
\hspace{-.1\textwidth}=&\int_{\gamma_{12}} d\sigma \int_{\gamma_{13}}d\sigma^{\prime}K_{\Delta_{1},\spin_{1}}(P_{1},U_{1};X_{\sigma})\mathbb{D}K_{\Delta_{2},\spin_{2}}(P_{2},U_{2};X_{\sigma})\mathbb{D}\mathbb{D}^{\prime}\bbprop_{\Delta,\spin}(X_{\sigma};X_{\sigma^{\prime}})K_{\Delta_{3},\spin_{3}}(P_{3},U_{3};X_{\sigma^{\prime}})\mathbb{D}^{\prime}K_{\Delta_{4},\spin_{4}}(P_{4},U_{4};X_{\sigma^{\prime}})\\
\hspace{-.1\textwidth}&\hspace{.25\textwidth}=\hat{\bbprop}_{\{\Delta_{i},\spin_{i}\};\Delta,\spin}^{IJ \; (GWD)}(P_{i},U_{i})
\,.
}

The immediate result of this equivalence is that the GWD presentation does indeed give the spinning conformal block. In particular the short distance behavior is  correct, since the boundary projector $\mathcal{P}_{\Delta}$ selects this behavior.

\subsection{Hybrid Presentation}\label{sec:hyb}

In the above sections, we have stressed how the geodesic Witten diagram presentation of the block, represented as an integral over two geodesics, and the shadow integral presentation of the block represented as an integral over the $d$ dimensional boundary, can be simply related using the split representation of the bulk-to-bulk propagator. Here we will outline a hybrid presentation of the block as a single geodesic integral. For simplicity of presentation, we will mostly discuss the harmonic function, $\Omega$, and make connection to the block itself at the end of the section.

The GWD presentation of the harmonic function can be written as in \eqref{geodesicblock}.
\es{geodesicharm}{
\hat{\Omega}_{\{\Delta_{i},\spin_{i}\};\Delta,\spin}^{IJ}(P_i)&=\int_{\gamma_{12}} d\sigma \int_{\gamma_{13}}d\sigma^{\prime}\langle\langle O_{1}(P_{1},\polvec_{1})O_{2}(P_{2},\polvec_{2})V^{I}(X_{\sigma})V^{J}(X_{\sigma^{\prime}})O_{3}(P_{3},\polvec_{3})O_{4}(P_{4},\polvec_{4})\rangle\rangle_{0}\\
&=\frac{d-2\D}{\ell!(h-1)_\ell}\int d\sigma d\sigma^{\prime}d^{d}P\left[\langle O_{1}(P_{1},\polvec_{1})O_{2}(P_{2},\polvec_{2})V^{I}(X_{\sigma})O_{\Delta,\spin}(P,D_{U})\rangle\right.\\
& \ \ \ \ \ \ \ \ \ \ \ \ \ \ \ \ \ \ \ \ \times \left.\langle\tilde{O}_{d-\Delta,\spin}(P,U)V^{J}(X_{\sigma^{\prime}})O_{3}(P_{3},\polvec_{3})O_{4}(P_{4},\polvec_{4})\rangle_{0}\right]\,.
} 
Here the double brackets, $\langle\langle\cdot\rangle\rangle_{0}$, in the first line indicate that we are using the bulk harmonic function, rather than the bulk-to-bulk propagator. In the second line we have used the split representation of the bulk harmonic function \eqref{splitJ}.

As mentioned above, if we wanted to make contact with the shadow formalism, we could perform the integrals over the bulk geodesics, producing boundary three point functions. Here, however, we will perform an integral only over the first geodesic, $\gamma_{12}$. This produces one of the three point structures, $\tilde{\mathcal{V}}^{I}$, from \eqref{struc}.
\es{hybharm}{
\hat{\Omega}_{\{\Delta_{i},\spin_{i}\};\Delta,\spin}^{IJ}(P_i)&=\frac{d-2\D}{\ell!(h-1)_\ell}\int d\sigma^{\prime}d^{d}P\left[\tilde{\mathcal{V}}^{I}(P_{1},\polvec_{1};P_{2},\polvec_{2};P,D_{U})\right.\\
& \ \ \ \ \ \ \ \ \ \ \ \ \ \ \ \ \ \ \ \ \times \left.\langle\tilde{O}_{d-\Delta,\spin}(P,U)V^{J}(X_{\sigma^{\prime}})O_{3}(P_{3},\polvec_{3})O_{4}(P_{4},\polvec_{4})\rangle_{0}\right]\,.
} 
We would now like to perform the $P$ integral over the boundary, to produce a one dimensional integral representation of the harmonic function. The salient piece of this integral takes the form,
\es{hybboundint}{
\int d^{d}P &\, \tilde{\mathcal{V}}^{I}(P_{1},\polvec_{1};P_{2},\polvec_{2};P,D_{U}) K_{d-\Delta,\spin}(P,U;X_{\sigma^{\prime}},W)\\
&\propto\int d^{d}P \,\frac{\tilde{\mathcal{V}}^{I}(P_{1},\polvec_{1};P_{2},\polvec_{2};P,D_{U})(U\cdot C(X_{\sigma^{\prime}},W)\cdot P)^{\spin}}{(-2P_{1}\cdot P)^{\frac{\tau+\tau_{12}}{2}}(-2P_{2}\cdot P)^{\frac{\tau-\tau_{12}}{2}}(-2P\cdot X_{\sigma^{\prime}})^{\frac{d-\Delta+\spin}{2}}}\,.
}
This integral, like the integral evaluated in Appendix~\ref{app:confint}, is an example of the conformal integrals computed in \cite{SimmonsDuffin:2012uy}. It can be evaluated in terms of sums of hypergeometric $_{2}F_{1}$ functions, and thus the harmonic function, \eqref{hybharm}, can be written as a one dimensional integral of sums of hypergeometric functions.\footnote{Note, a similar one dimensional integral representation is also provided directly by the shadow formalism. In the shadow presentation of the harmonic function, the $P$ integral over the boundary can be rewritten as an integral over three Feynman parameters. The integral over two of these parameters produces sums of hypergeometric functions, leaving one remaining integral. Here we have exchanged a Feynman parameter integral for a geodesic one.}

A similar statement may be made about the block, rather than the harmonic function, by acting with projectors, or equivalently by writing the bulk-to-bulk propagator as a sum over bulk harmonic functions (c.f. \cite{Costa:2014kfa} e.q. 95). 
\section{Discussion}\label{sec:discussion}

In this paper we have constructed geodesic Witten diagrams that compute CFT conformal blocks with spinning external and internal traceless symmetric tensor fields. To do this, we defined a new basis of CFT three-point functions and constructed their dual bulk vertices.  
The equivalence between the bulk geodesic Witten diagrams and boundary conformal blocks was most easily demonstrated by the use of the shadow-operator formalism: the shadow-operator representation of a CFT harmonic function directly translates to the split-representation of a bulk geodesic Witten diagram.

This highlights an advantage of bulk physics for computing conformal blocks. The shadow-operator methods of \cite{SimmonsDuffin:2012uy} give useful, compact expressions for spinning conformal blocks, but are complicated by the unpleasant necesssity of projecting out the unphysical shadow blocks. 
The corresponding bulk field, however, only contains the normalizable, propagating mode in AdS. 
This is equivalent to having already projected out the unphysical shadow. 
Thus, bulk GWD expressions for conformal blocks do not require us to act with a projector. 
For some, this may be a \emph{raison d'\^etre} for bulk physics, and GWDs provide a useful packaging of conformal blocks even when the CFT does not have a bulk dual. 

A spinning geodesic Witten diagram can also be viewed as the two point-function of two geodesic operators. In this way, our GWDs for spinning conformal blocks  allow an immediate derivation of a geodesic operator representation for the corresponding OPE blocks, generalizing the work of \cite{Czech:2016xec}. 

Another advantage of the GWD formalism is that they give relatively palatable two-dimensional integral expressions for spinning conformal blocks (as was also true of the scalar case \cite{Hijano:2015zsa}). 
As we have sketched in Section~\ref{sec:hyb}, one can do better still and write the spinning blocks as a finite sum of one dimensional integrals over a single geodesic. It would be interesting to examen these integrals in greater detail and see if they provide convenient presentations of the spinning conformal blocks.

Geodesic Witten diagrams may  illuminate  the question of the emergence of bulk geometry since 
the interactions of bulk fields with gravity are described in the CFT by the exchange of conformal blocks containing the stress tensor. 
A necessary precursor to understanding these exchanges is to determine the stress-tensor OPE coefficients from appropriate correlation functions. 

Lastly, the decomposition of CFT correlation functions into conformal blocks is fixed by kinematics and is well-understood as summing contributions from descendant operators. 
The corresponding decomposition of bulk physics into geodesic Witten diagrams must similarly be fixed by symmetry. On a case by case basis, it is of course possible to decompose the standard Witten diagrams in terms of geodesic Witten diagrams, thus yielding the bulk block decomposition, however it remains mysterious what the bulk structure that allows such a decomposition is in general.
It would be useful to understand this more directly.  

\section*{Acknowledgments}

We are grateful to
Aitor Lewkowycz and
Eric Perlmutter
for useful discussions and comments and
Alejandra Castro,
Eva Llabr\'es,
and Fernando Rejon-Barrera
for making us aware of their work and coordinating publication.
The research of ED was supported in part by the National Science Foundation under
grant NSF-PHY-1316699 and by the Stanford Institute for Theoretical Physics.
The research of DZF is partly supported by the US National Science Foundation grant NSF PHY-1620045 and as Templeton Visiting Professor at Stanford.   
JS is supported by the Natural Sciences and Engineering Research Council of Canada (NSERC) Banting Postdoctoral Fellowships program.

\appendix

\section{Propagators  in embedding space}\label{app:props}

In this Appendix, we present useful information on the bulk-boundary and bulk-bulk propagators 
for bulk tensor fields dual to CFT operators with scale dimension and spin $\D,~\ell$.  This material is mostly a recapitulation of \cite{Costa:2014kfa}.

The   bulk-boundary propagator satisfies the free homogeneous
field equation and  has vanishing divergence:
\be\lab{pieq}
(\nabla^2 -\D(\D-d)+\ell) K_{\D,\ell}(X,W;P,\polvec) =0\qquad\quad (K^A\nabla_A) \,K_{\D,\ell}(X,W;P,\polvec) =0.
\ee
(The divergence operator $(K^A\nabla_A)$ is defined in (19) of \cite{Costa:2014kfa}.) The explicit  embedding  space  polynomial   is
\bea\lab{pigseries}
K_{\D,\ell}(X,W;P,\polvec) &=& \bbnorm{\D}{\spin} \frac{[(-2P\cdot X)(W\cdot \polvec)+(2W\cdot   P)(\polvec\cdot X)]^\ell}{(-2P\cdot X)^{\D+\ell}}\\
\bbnorm{\D}{\spin} &=& \frac{(\ell+\D-1)\G(\D)}{2\pi^{h}(\D-1)\G(\D+1-h))}\,.
\eea
This is transverse and scales as $\l^{-\D}$ under $P\to \l P$.

The  bulk-bulk propagator  $\bbprop_{\D,\ell}(X,W; X',W')$  is also divergence free, and satisfies the wave equation with a $\d$-function source\footnote{ We ignore the additional local source  terms discussed in \cite{Costa:2014kfa}, because we need this propagator only for separated points.} 
\be
(\nabla^2 -\D(\D-d)+\ell)\bbprop_{\D,\ell}(X,W; X',W')= -\d(X,X')(W_1\cdot W_2)^\ell\,.
\ee
The mass is $M^2 = \D(\D-d)+\ell$.
This propagator has the representation
\bea
\bbprop_{\D,\ell}(X,W; X',W')&=&\sum_{k=0}^\ell (W\cdot W')^{\ell-k}((W\cdot X')(W'\cdot X))^k g_k(\bar u)\\
\bar u = -(1+X\cdot X') 
\eea
The variable $\bar u$ is the commonly used  AdS chordal  distance.  The polynomial above is not transverse, and  polarization vectors are stripped using the interior derivative operator $K_A$ defined in (12) of \cite{Costa:2014kfa}.

In the limit of large $\bar u$, the leading behavior of $g_k(\bar u)$ is the power law
\be
g_k(\bar u)\approx c_k \bar u^{-(\D+k)},
\ee
together with subsidiary powers $-(\D+k+j)$ suppressed by the positive integer $j$.  Therefore, under the monodromy $\bar u\to e^{-2\pi i}\bar u$,  all $g_k(\bar u)$ acquire the phase $e^{2\pi i \D}$.  The bulk propagator $\bbprop_{\D,\ell}(X,W; X',W')$ acquires the same phase, and falls as $\bar u^{-\D}$ for large $\bar u$.

The bulk-bulk propagator  approaches  the bulk-boundary propagator as the bulk point $X'$ goes to the boundary.  This limit is implemented by setting $X' = \lambda P + O(1/\l)$ to maintain $X'^2=-1$ and with $W'=\polvec$.  Then
\be
K_{\D,\ell}(X,W;P,\polvec) = \lim_{\l\to \infty} \l^\D \bbprop_{\D,\ell}(X,W; \l P+O(1/\l),\polvec).
\ee
We see that only the leading term $g_0(\bar u)$  of \reef{pigseries} contributes in this limit.

Given a bulk operator of dimension and spin $\D,~\ell$ we also consider its shadow  with dimension and spin $d-\D,~\ell$.  Its bulk propagator  $\bbprop_{d-\D,\ell}$ also satisfies (\ref{pieq}-\ref{pigseries}) with the same $M^2$.  
As  $\bar u\to\infty$, we see that $\bbprop_{d-\D,\ell}$ has monodromy phase $e^{2\pi i (d-\D)}$ and has leading power law $\bar u^{-(d-\D)}$.
The difference $\bbprop_{\D,\ell}- \bbprop_{d-\D,\ell}$ is the solution of a partial differential equation with no $\d$-function singularity. It is therefore called the bulk harmonic function and is precisely defined as\footnote{Note, this normalization differs slightly from that in \cite{Costa:2014kfa}.}
\be\lab{harm}
\O_{\D,\ell}(X,W; X',W')\equiv \bbprop_{\D,\ell}(X,W; X',W')- \bbprop_{d-\D,\ell}(X,W; X',W').
\ee

The harmonic function can be expressed  as a boundary integral of the bulk-boundary propagators of the direct and shadow operators via the split representation:
\be \lab{splitJ}
\O_{\D,\ell}(X,W; X',W')=\frac{d-2\D}{\ell!(h-1)_\ell}\int dP\,K_{\D,\ell}(X,W;P,D_\polvec)K_{d-\D,\ell}(X',W';P,\polvec)\,,
\ee
where $(h-1)_\ell$ is the Pochammer symbol, and $D_\polvec$ is the differential operator defined in \eqref{eq:free-bound-index}.

The bulk-boundary propagators of the direct and shadow operators are related by the boundary integral
(see (230) of \cite{Costa:2014kfa} )
\be\lab{dirshadJ}
\frac{1}{(2\D-d) C_{d-\D}} K_{d-\D,\ell}(X',W';P,\polvec)= \int dP'\frac{K_{\D,\ell}(X',W';P',D_\polvec') (H(\polvec,\polvec'))^\ell}{\ell!(h-1)_\ell(-2P\cdot P')^{d-\D}}\,.
\ee
Here $H(\polvec,\polvec')$ is the invariant of \eqref{Hdef} with $\polvec_1, P_1\to \polvec, P$ and $\polvec_2,P_2\to \polvec',\to P'$.  
The quantity $T_{\{A\}\{B\}}(P,P^{\prime})$ in \eqref{shadowint} is obtained from $(H(\polvec,\polvec'))^\ell/{\ell!(h-1)_\ell(-2P\cdot P')^{d-\D}}$ by stripping polarization vectors.


\section{Monodromy}\label{app:monod}
The monodromy situation we deal with in Secs.\ref{sec: shadow1},  \ref{sec:shadows-short}, is quite simple. We deal with functions of the form
\be
f(z) = z^\a H_\a(z) + z^\b H_\b(z),
\ee
where $H_\a(z),~H_\b(z)$ are holomorphic at $z=0$.
If we move $z$ around the branch point, i.e. we consider the monodromy  $z\to e^{2\pi i} z$,  we find
\be
e^{-2 \pi i\b} f(e^{2\pi i}z) = e^{2\pi i(\a-\b)} z^\a H_\a(z) +  z^\b H_\b(z). 
\ee
Then the component of $f(z)$  with monodromy phase $e^{2 \pi i\a}$  is
\be
z^\a H_\a(z)=\frac{f(z) - e^{-2\pi i\b}f(e^{2\pi i}z)}{1-e^{2\pi i(\a-\b)}}\,.
\ee
  
\paragraph{Example: the bulk scalar harmonic function $\O_\D(X,X')$.}
From Sec. 4 and App. C of \cite{Costa:2014kfa} we learn that the harmonic function \reef{harm} and bulk propagator $G_\D(X,X')$ of a scalar field with mass $m^2 =\D(\D-d)$ are related by 
\be\lab{omegaG}
\O_\D(X,X') = \frac{\D-d/2}{2\pi} \bigg(G_\D(X,X') - G_{d-\D}(X,X')\bigg)\,.
\ee
 The direct propagator is
\bea \lab{bulbulprop}
&&G_\D(X,X') =  \frac{\G(\D)}{2\pi^{ d/2}\G(\D+1 -d/2)}\frac{1}{(2\bar u)^\D} {}_2F_1(\D,\D-(d-1)/2\,;\, 2\D-d+1;-2/\bar u)\\
 &&(\Box_X -\D(\D-d)) G_\D(X,X')= - \d(X,X')\quad\quad \bar u = -(1+X\cdot X')= \frac{(x_0-x'_0)^2+(x_i-x'_i)^2}{2x_0x'_0},\nonumber
 \eea
where $\bar u$ is the usual chordal distance on AdS.  The shadow propagator is obtained from this by the replacement $\D\to d-\D$.

To extract the direct component of $\O_\D(X,X')$, we consider the monodromy at $\bar u = \infty,$ since the hypergeometric functions in \reef{omegaG} are holomorphic in a neighborhood of $\infty$.  To relate this to
the toy prototype above we set  $z=1/\bar u$ and define $f(z) = F(\bar u) = \O_\D(X,X').$  Then $F(\bar u)$ has the structure
\be \lab{Fstruc}
F(\bar u) = \frac{1}{\bar u^\D}F_\D(1/\bar u) + \frac{1}{\bar u^{d-\D}}F_{d-\D}(1/\bar u)
\ee
The monodromy $z\to e^{2\pi i} z$ is equivalent to $\bar u \to e^{-2\pi i} \bar u.$   The direct component of $F(\bar u)$ is 
\be\lab{harmdirect}
\bar u^{-\D} F_\D(1/\bar u)=\frac{F(\bar u) - e^{2\pi i(\D-d)} F(e^{-2\pi i} \bar u)}{1- e^{2\pi i(2\D-d)}}\,.
\ee

\section{Selected Derivations}\label{app:oddsends}
Here we collect two computations we felt would have disrupted the flow of the main text, these are the derivation of the relation used in \eqref{CasF} relating the Casimir and the bulk Laplacian, and \eqref{eq:shadowint} giving the shadow integral explicitly as a sum of direct and shadow blocks.
\subsection{Bulk Laplacian and Casimir}\label{app:cas}
For scalar bulk functions, the laplacian can be written as,
\es{scalap}{
\nabla^{2}f(X)&=G^{AB}\partial_{A}G_{B}^{B^{\prime}}\partial_{B^{\prime}}f(X)\\
&=G^{AB}\left[\eta_{AB}X^{B^{\prime}}+\delta_{A}^{B^{\prime}}X_{B}+G_{B}^{B^{\prime}}\partial_{A}\right]\partial_{B^{\prime}}f(X)\\
&=\left[(D-1)X\cdot\partial+G^{AB}\partial_{A}\partial_{B}\right]f(X)\,.
}
Note that this is simply related to the Casimir under bulk Lorentz rotations.
\es{scaLorCas}{
\mathcal{L}^{AB}\mathcal{L}_{AB}f(X)&=\left(X^{A}\partial^{B}-X^{B}\partial^{A}\right)\left(X_{A}\partial_{B}-X_{B}\partial_{A}\right)f(X)\\
&=-2\left[(D-1)X\cdot\partial+G^{AB}\partial_{A}\partial_{B} \right]\\
&=-2\nabla^{2}f(X)\,.
}
This relation can be generalized to operators with spin. For spinning operators,
\es{spinLap}{
\nabla^{2}f(X,W)&=\left[(D-1)X\cdot\partial+G^{AB}\partial_{A}\partial_{B}+W^{A}\frac{\partial}{\partial W^{A}}\right]f(X,W)\,,
}
while the Casimir gives,
\es{scaLorCas2}{
\mathcal{L}^{AB}\mathcal{L}_{AB}f(X,W)&=\left[\mathcal{L}^{(X)AB}\mathcal{L}^{(X)}_{AB}+\mathcal{L}^{(W)AB}\mathcal{L}^{(W)}_{AB}+2\mathcal{L}^{(X)AB}\mathcal{L}^{(W)}_{AB}\right]f(X,W)\,.
}
Each of these three terms give,
\es{eachCas}{
\mathcal{L}^{(X)AB}\mathcal{L}^{(X)}_{AB}f(X,W)&=-2\left(\nabla^{2}-W^{A}\frac{\partial}{\partial W^{A}}\right)f(X,W)\\
&=-2\left(\nabla^{2}-\spin\right)f(X,W)\\
\mathcal{L}^{(W)AB}\mathcal{L}^{(W)}_{AB}f(X,W)&=-2\left(W^{A}\frac{\partial}{\partial W^{A}}W^{B}\frac{\partial}{\partial W^{B}}+(D-2) W^{A}\frac{\partial}{\partial W^{A}}\right)f(X,W)\\
&=-2\spin\left(\spin+D-2\right)f(X,W)\\
\mathcal{L}^{(X)AB}\mathcal{L}^{(W)}_{AB}f(X,W)&=-2X^{A}W^{B}\frac{\partial}{\partial W^{A}}\partial_{B}f(X,W)\\
&=2\spin f(X,W)\,.
}
Putting these together,
\es{Caseq}{
\mathcal{L}^{AB}\mathcal{L}_{AB}f(X,W)&=-2\left(\nabla^{2}+\spin(d+\spin-1)\right)f(X,W)\,.
}
\subsection{Conformal Integrals}\label{app:confint}
Here we review the scalar shadow integral which produces a sum of direct and shadow conformal blocks. The result is equation \eqref{eq:shadowint} in the main text.
We want to evaluate the four-point integral,
\es{shadowInt}{
\hspace{-.13\textwidth}\Omega_{\Delta_{i};\Delta}(P_{i})&=\int d^{d}P\langle O_{1}(P_{1})O_{2}(P_{2})O(P)\rangle\langle\tilde{O}(P)O_{3}(P_{3})O_{4}(P_{4})\rangle\\
\hspace{-.13\textwidth}\Omega_{\Delta_{i};\Delta}(P_{i})&=\frac{1}{P_{12}^{\frac{\Delta_{1}+\Delta_{2}-\Delta}{2}}P_{34}^{\frac{\Delta_{3}+\Delta_{4}+\Delta-d}{2}}} \\
&\quad\quad \times \int d^{d}P\frac{1}{(-2P_{1}\cdot P)^{\frac{\Delta_{1}+\Delta-\Delta_{2}}{2}}(-2P_{2}\cdot P)^{\frac{\Delta_{2}+\Delta-\Delta_{1}}{2}}(-2P_{3}\cdot P)^{\frac{\Delta_{3}+d-\Delta-\Delta_{4}}{2}}(-2P_{4}\cdot P)^{\frac{\Delta_{4}+d-\Delta-\Delta_{3}}{2}}}\,.
}
Let's start by evaluating the auxiliary integral,
\es{gen4pt}{
I_{4}&=\int d^{d}P\frac{1}{(-2P_{1}\cdot P)^{a_{1}}(-2P_{2}\cdot P)^{a_{2}}(-2P_{3}\cdot P)^{a_{3}}(-2P_{4}\cdot P)^{a_{4}}}\\
&=\frac{\Gamma(d)}{\prod_{i=1}^{4}\Gamma(a_{i})}\int d^{d}P\int \prod_{i=2}^{4}\frac{d\alpha_{i}}{\alpha_{i}}\frac{\prod_{i=2}^{4}\alpha_{i}^{a_{i}}}{\left(-2P\cdot(P_{1}+\sum_{i=2}^{4}\alpha_{i}P_{i})\right)^{d}}\\
&=\frac{\pi^{d/2}\Gamma(d/2)}{\prod_{i=1}^{4}\Gamma(a_{i})}\int \prod_{i=2}^{4}\frac{d\alpha_{i}}{\alpha_{i}}\frac{\prod_{i=2}^{4}\alpha_{i}^{a_{i}}}{(\alpha_{2}P_{12}+\alpha_{3}P_{13}+\alpha_{4}P_{14}+\alpha_{2}\alpha_{3}P_{23}+\alpha_{2}\alpha_{4}P_{24}+\alpha_{3}\alpha_{4}P_{34})^{d/2}}\\
&=\frac{\pi^{d/2}\Gamma(a_{4})\Gamma(d/2-a_{4})}{\prod_{i=1}^{4}\Gamma(a_{i})}\int \prod_{i=2}^{3}\frac{d\alpha_{i}}{\alpha_{i}}\frac{\prod_{i=2}^{3}\alpha_{i}^{a_{i}}}{(\alpha_{2}P_{12}+\alpha_{3}P_{13}+\alpha_{2}\alpha_{3}P_{23})^{d/2-a_{4}}(P_{14}+\alpha_{2}P_{24}+\alpha_{3}P_{34})^{a_{4}}}\\
&=\frac{\pi^{d/2}\Gamma(a_{4})\Gamma(d/2-a_{4})}{\prod_{i=1}^{4}\Gamma(a_{i})}\frac{P_{13}^{a_4-\frac{d}{2}} P_{14}^{a_2+a_3-\frac{d}{2}}}{P_{24}^{a_2}P_{34}^{a_3+a_4-\frac{d}{2}}}\int \prod_{i=2}^{3}\frac{d\hat{\alpha}_{i}}{\hat{\alpha}_{i}}\frac{\prod_{i=2}^{3}\hat{\alpha}_{i}^{a_{i}}}{(\hat{\alpha}_{2}u+\hat{\alpha}_{3}+\hat{\alpha}_{2}\hat{\alpha}_{3}v)^{d/2-a_{4}}(1+\hat{\alpha}_{2}+\hat{\alpha}_{3})^{a_{4}}}\\
&=\frac{\pi^{d/2}\Gamma(a_{4})\Gamma(d/2-a_{4})}{\prod_{i=1}^{4}\Gamma(a_{i})}\frac{P_{13}^{a_4-\frac{d}{2}} P_{14}^{a_2+a_3-\frac{d}{2}}}{P_{24}^{a_2}P_{34}^{a_3+a_4-\frac{d}{2}}}\\
&\ \ \ \times \int d\hat{\alpha}_{2}\left(f\left(a_{2},a_{3},a_{4};\hat{\alpha}_{2};u,v)+u^{a_{3}+a_{4}-\frac{d}{2}}f(a_{2}+a_{3}+a_{4}-\frac{d}{2},\frac{d}{2}-a_{4},\frac{d}{2}-a_{3};\hat{\alpha}_{2};u,v\right)\right)\,,
}

where, $u=\frac{P_{12}P_{34}}{P_{13}P_{24}}$, $v=\frac{P_{14}P_{23}}{P_{13}P_{24}}$, and
\es{intnote}{
f(a_{2},a_{3},a_{4};\hat{\alpha}_{2};u,v)&=\frac{\pi ^{d/2} \Gamma \left(\frac{d}{2}-a_3\right) \Gamma \left(\frac{d}{2}-a_4\right) \Gamma \left(-\frac{d}{2}+a_3+a_4\right)}{\Gamma (d)}\alpha _2^{a_2-1} \left(\alpha _2+1\right){}^{a_3-\frac{d}{2}} \left(\alpha _2 v+1\right){}^{a_4-\frac{d}{2}}\\
&\ \ \times \, _2F_1\left(\frac{d}{2}-a_3,\frac{d}{2}-a_4;-a_3-a_4+\frac{d}{2}+1;\frac{\alpha _2 u}{\left(\alpha _2+1\right) \left(\alpha _2 v+1\right)}\right)\,.
}

Plugging this into \eqref{shadowInt}, the two terms in the last line of \eqref{gen4pt} become the direct and shadow block respectively,
\es{blockdecomp2}{
\Omega_{\Delta_{i};\Delta}(P_{i})&=\left(\frac{P_{14}}{P_{13}}\right)^{\frac{\Delta_{3}-\Delta_{4}}{2}}\left(\frac{P_{24}}{P_{14}}\right)^{\frac{\Delta_{1}-\Delta_{2}}{2}}\frac{\mathcal{N}_{\{\Delta_{i}\}}}{{P_{12}^{\frac{\Delta_{1}+\Delta_{2}}{2}}P_{34}^{\frac{\Delta_{3}+\Delta_{4}}{2}}}}\left(\bbprop_{\Delta_{i};\Delta}(u,v)-\bbprop_{\Delta_{i};d-\Delta}(u,v)\right)\,,
}
with $\mathcal{N}_{\{\Delta_{i}\}}=\frac{1}{\Delta-\frac{d}{2}}\frac{\Gamma \left(\frac{1}{2} \left(\Delta +\Delta _3-\Delta _4\right)\right) \Gamma \left(\frac{1}{2} \left(\Delta -\Delta _3+\Delta _4\right)\right)}{\Gamma \left(\frac{1}{2} \left(\Delta +\Delta _1-\Delta _2\right)\right) \Gamma \left(\frac{1}{2} \left(\Delta -\Delta _1+\Delta _2\right)\right)}$, and
\es{blockdefapp}{
\bbprop_{\Delta_{i};\Delta}(u,v)&=\frac{(\Delta-d/2)\Gamma \left(\frac{d}{2}-\Delta \right)}{\Gamma \left(\frac{1}{2} \left(d-\Delta +\Delta _3-\Delta _4\right)\right) \Gamma \left(\frac{1}{2} \left(d-\Delta -\Delta _3+\Delta _4\right)\right)}\\
&\times u^{\frac{\Delta}{2}}\int d\alpha_{2}\left(\alpha _2+1\right){}^{\frac{1}{2} \left(-\Delta +\Delta _3-\Delta _4\right)} \left(\alpha _2 v+1\right){}^{\frac{1}{2} \left(-\Delta -\Delta _3+\Delta _4\right)} \alpha _2^{\frac{1}{2} (\Delta -\Delta_{1}+\Delta_{2}-2)}\\
&\ \ \ \ \ \ \ \ \ \times \, _2F_1\left(\frac{\left(\Delta +\Delta _3-\Delta _4\right)}{2},\frac{\left(\Delta -\Delta _3+\Delta _4\right)}{2};-\frac{d}{2}+\Delta +1;\frac{\alpha _2 u}{\left(\alpha _2+1\right) \left(\alpha _2 v+1\right)}\right)\,.
}
Up to the overall normalization, This integral expression matches previous expressions in the literature (for example, \cite{Hijano:2015zsa} equation (2.10)).

\bibliography{refs}{}

\providecommand{\href}[2]{#2}\begingroup\raggedright\begin{thebibliography}{10}

\bibitem{Hijano:2015zsa}
E.~Hijano, P.~Kraus, E.~Perlmutter, and R.~Snively, {\it {Witten Diagrams
  Revisited: The AdS Geometry of Conformal Blocks}},  {\em JHEP} {\bf 01}
  (2016) 146, [\href{http://xxx.lanl.gov/abs/1508.00501}{{\tt 1508.00501}}].

\bibitem{Rattazzi:2008pe}
R.~Rattazzi, V.~S. Rychkov, E.~Tonni, and A.~Vichi, {\it {Bounding scalar
  operator dimensions in 4D CFT}},  {\em JHEP} {\bf 12} (2008) 031,
  [\href{http://xxx.lanl.gov/abs/0807.0004}{{\tt 0807.0004}}].

\bibitem{Rychkov:2009ij}
V.~S. Rychkov and A.~Vichi, {\it {Universal Constraints on Conformal Operator
  Dimensions}},  {\em Phys. Rev.} {\bf D80} (2009) 045006,
  [\href{http://xxx.lanl.gov/abs/0905.2211}{{\tt 0905.2211}}].

\bibitem{Caracciolo:2009bx}
F.~Caracciolo and V.~S. Rychkov, {\it {Rigorous Limits on the Interaction
  Strength in Quantum Field Theory}},  {\em Phys. Rev.} {\bf D81} (2010)
  085037, [\href{http://xxx.lanl.gov/abs/0912.2726}{{\tt 0912.2726}}].

\bibitem{Poland:2010wg}
D.~Poland and D.~Simmons-Duffin, {\it {Bounds on 4D Conformal and
  Superconformal Field Theories}},  {\em JHEP} {\bf 05} (2011) 017,
  [\href{http://xxx.lanl.gov/abs/1009.2087}{{\tt 1009.2087}}].

\bibitem{Rattazzi:2010gj}
R.~Rattazzi, S.~Rychkov, and A.~Vichi, {\it {Central Charge Bounds in 4D
  Conformal Field Theory}},  {\em Phys. Rev.} {\bf D83} (2011) 046011,
  [\href{http://xxx.lanl.gov/abs/1009.2725}{{\tt 1009.2725}}].

\bibitem{Rattazzi:2010yc}
R.~Rattazzi, S.~Rychkov, and A.~Vichi, {\it {Bounds in 4D Conformal Field
  Theories with Global Symmetry}},  {\em J. Phys.} {\bf A44} (2011) 035402,
  [\href{http://xxx.lanl.gov/abs/1009.5985}{{\tt 1009.5985}}].

\bibitem{Vichi:2011ux}
A.~Vichi, {\it {Improved bounds for CFT's with global symmetries}},  {\em JHEP}
  {\bf 01} (2012) 162, [\href{http://xxx.lanl.gov/abs/1106.4037}{{\tt
  1106.4037}}].

\bibitem{Poland:2011ey}
D.~Poland, D.~Simmons-Duffin, and A.~Vichi, {\it {Carving Out the Space of 4D
  CFTs}},  {\em JHEP} {\bf 05} (2012) 110,
  [\href{http://xxx.lanl.gov/abs/1109.5176}{{\tt 1109.5176}}].

\bibitem{ElShowk:2012ht}
S.~El-Showk, M.~F. Paulos, D.~Poland, S.~Rychkov, D.~Simmons-Duffin, and
  A.~Vichi, {\it {Solving the 3D Ising Model with the Conformal Bootstrap}},
  {\em Phys. Rev.} {\bf D86} (2012) 025022,
  [\href{http://xxx.lanl.gov/abs/1203.6064}{{\tt 1203.6064}}].

\bibitem{Mack:1969rr}
G.~Mack and A.~Salam, {\it {Finite component field representations of the
  conformal group}},  {\em Annals Phys.} {\bf 53} (1969) 174--202.

\bibitem{Ferrara:1972xe}
S.~Ferrara and G.~Parisi, {\it {Conformal covariant correlation functions}},
  {\em Nucl. Phys.} {\bf B42} (1972) 281--290.

\bibitem{Ferrara:1972ay}
S.~Ferrara, A.~F. Grillo, and G.~Parisi, {\it {Nonequivalence between conformal
  covariant wilson expansion in euclidean and minkowski space}},  {\em Lett.
  Nuovo Cim.} {\bf 5S2} (1972) 147--151. [Lett. Nuovo Cim.5,147(1972)].

\bibitem{Ferrara:1972uq}
S.~Ferrara, A.~F. Grillo, G.~Parisi, and R.~Gatto, {\it {The shadow operator
  formalism for conformal algebra. vacuum expectation values and operator
  products}},  {\em Lett. Nuovo Cim.} {\bf 4S2} (1972) 115--120. [Lett. Nuovo
  Cim.4,115(1972)].

\bibitem{Ferrara:1973vz}
S.~Ferrara, A.~F. Grillo, G.~Parisi, and R.~Gatto, {\it {Covariant expansion of
  the conformal four-point function}},  {\em Nucl. Phys.} {\bf B49} (1972)
  77--98. [Erratum: Nucl. Phys.B53,643(1973)].

\bibitem{Ferrara:1973eg}
S.~Ferrara, R.~Gatto, and A.~F. Grillo, {\it {Conformal algebra in space-time
  and operator product expansion}},  {\em Springer Tracts Mod. Phys.} {\bf 67}
  (1973) 1--64.

\bibitem{Polyakov:1974gs}
A.~M. Polyakov, {\it {Nonhamiltonian approach to conformal quantum field
  theory}},  {\em Zh. Eksp. Teor. Fiz.} {\bf 66} (1974) 23--42.

\bibitem{Rychkov:2016iqz}
S.~Rychkov, {\em {EPFL Lectures on Conformal Field Theory in D>= 3
  Dimensions}}.
\newblock SpringerBriefs in Physics. 2016.

\bibitem{Simmons-Duffin:2016gjk}
D.~Simmons-Duffin, {\it {TASI Lectures on the Conformal Bootstrap}},
  \href{http://xxx.lanl.gov/abs/1602.07982}{{\tt 1602.07982}}.

\bibitem{Penedones:2016voo}
J.~Penedones, {\it {TASI lectures on AdS/CFT}},
  \href{http://xxx.lanl.gov/abs/1608.04948}{{\tt 1608.04948}}.

\bibitem{Hijano:2015rla}
E.~Hijano, P.~Kraus, and R.~Snively, {\it {Worldline approach to semi-classical
  conformal blocks}},  {\em JHEP} {\bf 07} (2015) 131,
  [\href{http://xxx.lanl.gov/abs/1501.02260}{{\tt 1501.02260}}].

\bibitem{Hijano:2015qja}
E.~Hijano, P.~Kraus, E.~Perlmutter, and R.~Snively, {\it {Semiclassical
  Virasoro blocks from AdS$_{3}$ gravity}},  {\em JHEP} {\bf 12} (2015) 077,
  [\href{http://xxx.lanl.gov/abs/1508.04987}{{\tt 1508.04987}}].

\bibitem{Nishida:2016vds}
M.~Nishida and K.~Tamaoka, {\it {Geodesic Witten diagrams with an external
  spinning field}},  \href{http://xxx.lanl.gov/abs/1609.04563}{{\tt
  1609.04563}}.

\bibitem{Costa:2011mg}
M.~S. Costa, J.~Penedones, D.~Poland, and S.~Rychkov, {\it {Spinning Conformal
  Correlators}},  {\em JHEP} {\bf 11} (2011) 071,
  [\href{http://xxx.lanl.gov/abs/1107.3554}{{\tt 1107.3554}}].

\bibitem{Costa:2014kfa}
M.~S. Costa, V.~Gonçalves, and J.~Penedones, {\it {Spinning AdS Propagators}},
  {\em JHEP} {\bf 09} (2014) 064,
  [\href{http://xxx.lanl.gov/abs/1404.5625}{{\tt 1404.5625}}].

\bibitem{SimmonsDuffin:2012uy}
D.~Simmons-Duffin, {\it {Projectors, Shadows, and Conformal Blocks}},  {\em
  JHEP} {\bf 04} (2014) 146, [\href{http://xxx.lanl.gov/abs/1204.3894}{{\tt
  1204.3894}}].

\bibitem{Dirac:1936fq}
P.~A.~M. Dirac, {\it {Wave equations in conformal space}},  {\em Annals Math.}
  {\bf 37} (1936) 429--442.

\bibitem{Boulware:1970ty}
D.~G. Boulware, L.~S. Brown, and R.~D. Peccei, {\it {Deep-inelastic
  electroproduction and conformal symmetry}},  {\em Phys. Rev.} {\bf D2} (1970)
  293--298.

\bibitem{Weinberg:2010fx}
S.~Weinberg, {\it {Six-dimensional Methods for Four-dimensional Conformal Field
  Theories}},  {\em Phys. Rev.} {\bf D82} (2010) 045031,
  [\href{http://xxx.lanl.gov/abs/1006.3480}{{\tt 1006.3480}}].

\bibitem{Costa:2011dw}
M.~S. Costa, J.~Penedones, D.~Poland, and S.~Rychkov, {\it {Spinning Conformal
  Blocks}},  {\em JHEP} {\bf 11} (2011) 154,
  [\href{http://xxx.lanl.gov/abs/1109.6321}{{\tt 1109.6321}}].

\bibitem{Alday:2013cwa}
L.~F. Alday and A.~Bissi, {\it {Higher-spin correlators}},  {\em JHEP} {\bf 10}
  (2013) 202, [\href{http://xxx.lanl.gov/abs/1305.4604}{{\tt 1305.4604}}].

\bibitem{Costa:2014rya}
M.~S. Costa and T.~Hansen, {\it {Conformal correlators of mixed-symmetry
  tensors}},  {\em JHEP} {\bf 02} (2015) 151,
  [\href{http://xxx.lanl.gov/abs/1411.7351}{{\tt 1411.7351}}].

\bibitem{Penedones:2015aga}
J.~Penedones, E.~Trevisani, and M.~Yamazaki, {\it {Recursion Relations for
  Conformal Blocks}},  {\em JHEP} {\bf 09} (2016) 070,
  [\href{http://xxx.lanl.gov/abs/1509.00428}{{\tt 1509.00428}}].

\bibitem{Rejon-Barrera:2015bpa}
F.~Rejon-Barrera and D.~Robbins, {\it {Scalar-Vector Bootstrap}},  {\em JHEP}
  {\bf 01} (2016) 139, [\href{http://xxx.lanl.gov/abs/1508.02676}{{\tt
  1508.02676}}].

\bibitem{Iliesiu:2015qra}
L.~Iliesiu, F.~Kos, D.~Poland, S.~S. Pufu, D.~Simmons-Duffin, and R.~Yacoby,
  {\it {Bootstrapping 3D Fermions}},  {\em JHEP} {\bf 03} (2016) 120,
  [\href{http://xxx.lanl.gov/abs/1508.00012}{{\tt 1508.00012}}].

\bibitem{Iliesiu:2015akf}
L.~Iliesiu, F.~Kos, D.~Poland, S.~S. Pufu, D.~Simmons-Duffin, and R.~Yacoby,
  {\it {Fermion-Scalar Conformal Blocks}},  {\em JHEP} {\bf 04} (2016) 074,
  [\href{http://xxx.lanl.gov/abs/1511.01497}{{\tt 1511.01497}}].

\bibitem{Echeverri:2015rwa}
A.~Castedo~Echeverri, E.~Elkhidir, D.~Karateev, and M.~Serone, {\it
  {Deconstructing Conformal Blocks in 4D CFT}},  {\em JHEP} {\bf 08} (2015)
  101, [\href{http://xxx.lanl.gov/abs/1505.03750}{{\tt 1505.03750}}].

\bibitem{Costa:2016hju}
M.~S. Costa, T.~Hansen, J.~Penedones, and E.~Trevisani, {\it {Projectors and
  seed conformal blocks for traceless mixed-symmetry tensors}},  {\em JHEP}
  {\bf 07} (2016) 018, [\href{http://xxx.lanl.gov/abs/1603.05551}{{\tt
  1603.05551}}].

\bibitem{Costa:2016xah}
M.~S. Costa, T.~Hansen, J.~Penedones, and E.~Trevisani, {\it {Radial expansion
  for spinning conformal blocks}},  {\em JHEP} {\bf 07} (2016) 057,
  [\href{http://xxx.lanl.gov/abs/1603.05552}{{\tt 1603.05552}}].

\bibitem{Echeverri:2016dun}
A.~Castedo~Echeverri, E.~Elkhidir, D.~Karateev, and M.~Serone, {\it {Seed
  Conformal Blocks in 4D CFT}},  {\em JHEP} {\bf 02} (2016) 183,
  [\href{http://xxx.lanl.gov/abs/1601.05325}{{\tt 1601.05325}}].

\bibitem{Schomerus:2016epl}
V.~Schomerus, E.~Sobko, and M.~Isachenkov, {\it {Harmony of Spinning Conformal
  Blocks}},  \href{http://xxx.lanl.gov/abs/1612.02479}{{\tt 1612.02479}}.

\bibitem{Fortin:2016dlj}
J.-F. Fortin and W.~Skiba, {\it {Conformal Differential Operator in Embedding
  Space and its Applications}},  \href{http://xxx.lanl.gov/abs/1612.08672}{{\tt
  1612.08672}}.

\bibitem{CastroEtAl}
A.~Castro, E.~Llabr\'es, and F.~Rejon-Barrera {\em To Appear}.

\bibitem{Dolan:2000ut}
F.~A. Dolan and H.~Osborn, {\it {Conformal four point functions and the
  operator product expansion}},  {\em Nucl. Phys.} {\bf B599} (2001) 459--496,
  [\href{http://xxx.lanl.gov/abs/hep-th/0011040}{{\tt hep-th/0011040}}].

\bibitem{Dolan:2003hv}
F.~A. Dolan and H.~Osborn, {\it {Conformal partial waves and the operator
  product expansion}},  {\em Nucl. Phys.} {\bf B678} (2004) 491--507,
  [\href{http://xxx.lanl.gov/abs/hep-th/0309180}{{\tt hep-th/0309180}}].

\bibitem{Freedman:1998tz}
D.~Z. Freedman, S.~D. Mathur, A.~Matusis, and L.~Rastelli, {\it {Correlation
  functions in the CFT(d) / AdS(d+1) correspondence}},  {\em Nucl. Phys.} {\bf
  B546} (1999) 96--118, [\href{http://xxx.lanl.gov/abs/hep-th/9804058}{{\tt
  hep-th/9804058}}].

\bibitem{Sleight:2016dba}
C.~Sleight and M.~Taronna, {\it {Higher Spin Interactions from Conformal Field
  Theory: The Complete Cubic Couplings}},  {\em Phys. Rev. Lett.} {\bf 116}
  (2016), no.~18 181602, [\href{http://xxx.lanl.gov/abs/1603.00022}{{\tt
  1603.00022}}].

\bibitem{Pilch:1984xx}
K.~Pilch and A.~N. Schellekens, {\it {Formulae for the Eigenvalues of the
  Laplacian on Tensor Harmonics on Symmetric Coset Spaces}},  {\em J. Math.
  Phys.} {\bf 25} (1984) 3455.

\bibitem{Czech:2016xec}
B.~Czech, L.~Lamprou, S.~McCandlish, B.~Mosk, and J.~Sully, {\it {A
  Stereoscopic Look into the Bulk}},  {\em JHEP} {\bf 07} (2016) 129,
  [\href{http://xxx.lanl.gov/abs/1604.03110}{{\tt 1604.03110}}].

\end{thebibliography}\endgroup
\bibliographystyle{jhep}

\end{document}